\documentclass[a4paper,12pt]{article}
\usepackage{amssymb,amsmath,cite}

\def\beq{\begin{equation}}
\def\eeq{\end{equation}}
\def\bea{\begin{eqnarray*}}
\def\eea{\end{eqnarray*}}
\def\nn{\nonumber}

\def\bra#1{\left\langle #1\right|}
\def\ket#1{\left| #1\right\rangle}
\def\bracket#1#2{\left\langle #1 | #2 \right\rangle}

\def\N0{ {\mathbb Z}_{+} }
\def\v0{ |d,r) }

\def\proof#1{{\bf Proof:} #1 $\blacksquare$ \medskip}

\newtheorem{lemma}{Lemma}
\newtheorem{prop}[lemma]{Proposition}
\newtheorem{thm}[lemma]{Theorem}
\newtheorem{cor}[lemma]{Corollary}
\begin{document}

\begin{center}
{\large\bf Invariants and reduced matrix elements associated with the Lie superalgebra $gl(m|n)$}\\
~~\\

{\large Mark D. Gould, Phillip S. Isaac and Jason L. Werry}\\
~~\\

School of Mathematics and Physics, The University of Queensland, St Lucia QLD 4072, Australia.
\end{center}

\begin{abstract}
We construct explicit formulae for the eigenvalues of certain invariants of the Lie
superalgebra $gl(m|n)$ using characteristic identities. We discuss how such eigenvalues are
related to reduced Wigner coefficients and the reduced matrix elements of generators, and thus provide a first step to a
new algebraic derivation of matrix element formulae for all generators of the algebra.
\end{abstract}

%

\section{Introduction}

The theory of basic classical Lie superalgebras was extensively developed in the late 1970s by
Kac \cite{Kac1977,Kac1978} and Scheunert et al. \cite{NaSch1976,SchNaRit1976,Sch1979},
motivated not only from the mathematical viewpoint of having a wondrous generalisation of the well
established theory of Lie algebras, but also by progress at the time in elementary
particle physics and generalised Fermi-Bose statistics
\cite{Ram1971,NevSchw1971,VolAk1973,WessZum1974,SalStrath1974,Scherk1975,FayFer1977}. We
direct the reader to an informative review of physical 
applications of Lie superalgebras that were known at the time
\cite{CorNeSt1975}. In more recent times, Lie superalgebras continue to be of pure mathematical interest 
(see, for example, the book by Musson \cite{Musson2012}), and lie at the heart of
many applications -- to give some key examples, they appear as symmetry algebras in ${\cal N}=4$ super
Yang-Mills theory \cite{BeiStau2003,Mina2012} and other supersymmetric integrable models
\cite{GalMar2004,EssFraSal2005,ZhYaZh2006,RagSat2007,FraMar2011}, underly logarithmic conformal field theories
\cite{SchomSal2006,Ridout2009} and play a role in systems combining parafermions and
parabosons \cite{StoiVan2008,LievStoiVan2008,StoiVan2010}. Undeniably, Lie superalgebras
have made their way into the mainstream of modern mathematical physics.

In most applications, it is important to have a well-developed representation theory of
the symmetry algebras involved. One fundamental question of representation theory is how
to provide explicit formulae for matrix elements of generators. For
Lie algebras, such a construction was unknown until the work of Gelfand and Tsetlin
\cite{GT1950,GT1950b} where formulae for matrix elements of simple generators for the
general linear and orthogonal Lie algebras were obtained. They moreover introduced combinatorial presentations of
the basis vectors, now commonly referred to as Gelfand-Tsetlin (GT) patterns. Partially
motivated by a desire to understand the results of these two brief articles, Baird and
Biedenharn \cite{BB1963} developed pattern calculus techniques in order to derive and extend the
remarkable formulae of Gelfand and Tsetlin. Many works followed (for a very readable
account, see the review article by Molev \cite{Molev2006} and references therein),
some of which are of particular interest to our present investigation,
including presentations of matrix element formulae
\cite{Palev1987,Palev1989,TolIstSmi1986,LievStoiVan2008,StoiVan2010,Molev2011} and branching rules
\cite{KamKyPal1989,GouBraHug1989,GouJarBra1990,PalStoi1990} for certain classes of
representations of a variety of Lie superalgebras. 

There is a body of literature from the 1970s and 1980s, 
that was devoted to determining characteristic (polynomial) identities satisfied by
generators of Lie algebras \cite{Green1971,BraGre1971,OBCantCar1977,Gould1985}.  
Curiously, such characteristic identities were noticed by Dirac as early as 1936
\cite{Dirac1936}, and their usefulness observed by Baird and Biedenharn
\cite{BB1964b}. 
Of particular note are the applications of these
characteristic identities 
to the derivation of reduced matrix elements \cite{Gould1978}, raising and
lowering generators \cite{Gould1980} and matrix elements \cite{Gould1981,Gould1981b}, even
in the context of infinite dimensional irreducible representations for semisimple Lie
algebras \cite{Gould1984,Kostant1975}.

A great deal is already known about Casimir invariants of Lie superalgebras and their
eigenvalues on irreducible representations \cite{Bin1983,GreJar1983,Sch1983,LinZha1993}.
In the current article, we seek to construct invariants related to reduced matrix elements
and reduced Wigner coefficients, in a similar vein to the treatment of classical Lie
algebras found in \cite{Gould1978,Gould1986,Gould1986b}. In order to determine eigenvalues
of these invariants on the irreducible representations, one could attempt to express them directly 
in terms of the Casimir invariants
(c.f. the work of Green \cite{Green1971} for classical Lie algebras). To our knowledge,
such an approach has not been attempted for Lie superalgebras, possibly for good reason. An alternative
strategy, as presented in the current article, makes use of characteristic identities, and
an important family of elements known as tensor operators.

Tensor operators play an important role in our work, especially since they
serve as intertwining operators (see Section \ref{to} for a more comprehensive discussion). 
Many textbooks on quantum mechanics present a treatment of tensor operators in the context
of $su(2)$ and the Wigner-Eckart theorem (see, for example, the book by Hannabuss
\cite{Hannabuss1997}).
The fact that tensor operators constitute intertwining operators in
more general cases such as Lie algebras other than $su(2)$
\cite{BB1964,LouBei1970,Gould1978,Gould1980,Gould1981,Gould1981b}, quantum groups
\cite{ZhGoBr1991} and Hopf algebras \cite{RitSch1992,Mozr2005}, 
allows many of the standard results for $su(2)$ to be extended. 
Discussions of tensor operators associated with Lie superalgebras have been presented for some special cases in
\cite{PaisRitt1975,Mezin1977,Mozr2004}, and it may seem at first that the situation is not so straightforward
in the general case. We seek to clarify this, and in doing so, explain how the eigenvalues of the
invariants that we construct are a first step in obtaining matrix element formulae for the
generators of $gl(m|n)$.

Of particular relevance to the current article is the seminal work of Jarvis and Green
\cite{JarGre1979} where characteristic identities were developed for the general linear,
special linear and orthosymplectic Lie superalgebras. Other works along these lines
include the development of characteristic identities associated to the so called ``strange'' Lie
superalgebras \cite{JarMur1983} and simple Lie superalgebras \cite{Gould1987}. 
More recently, techniques involving characteristic identities have been used to study the
representation theory of certain polynomial deformations of Lie superalgberas
\cite{JaRuYa2011}.
The current article is concerned with generalising the techniques
employing characteristic identities satisfied by generators of the Lie superalgebra
$gl(m|n)$, specifically to determining eigenvalues of invariants associated with tensor
operators. These invariants are of interest since their eigenvalues correspond to the squared reduced matrix
elements of the generators. 

The current article has two main goals: 
\begin{enumerate}
\item To highlight the effectiveness of the characteristic identities and the shift vector
formalism in determining eigenvalues of certain invariants related to reduced matrix
elements and reduced Wigner coefficients,
by generalising known methods to the case of the Lie superalgebra $gl(m|n)$; 
\item To make a first step in providing the details of the derivation of matrix element
formulae for all $gl(m|n)$ generators on irreducible representations.
\end{enumerate}
This second goal is largely in the spirit of Baird and Biedenharn \cite{BB1963}, ultimately aimed at
understanding the derivation of the matrix element formula, the focus being on the
means by which the formulae are derived, and will be the subject of future work. 

It is worth noting that our results have been obtained for any irreducible representation
of $gl(m|n)$, without any reference to unitary irreducible representations or their
classification \cite{ZhaGou1990}, for all generators (i.e. not just the simple ones), and
without dependence on the precise branching rules. Moreover, our approach unifies and
consolidates previous independent work of Palev \cite{Palev1987,Palev1989}, Stoilova and
Van der Jeugt \cite{StoiVan2010} and Molev \cite{Molev2011} (see also Tolstoy et al.
\cite{TolIstSmi1986}) into one unifying framework.

The article is organised as follows. Section \ref{pp} introduces the basic notations used
in the paper, and provides some basic constructions for Section
\ref{ci}, which establishes the form of the characteristic identities used throughout.
Section \ref{to} discusses tensor operators in a graded context, paving the way for
Section \ref{vosc} which introduces some of the main objects of our study -- vector
operators and their shift components. Section \ref{br} looks at the branching rules
associated with the subalgebra inclusion $gl(m|n+1)\supset gl(m|n)$ and establishes
necessary conditions in the form of betweenness
conditions. The key result of Section \ref{br} is given in Theorem \ref{TheoremBR}. 
It turns out that our approach does not require {\em precise} branching rules, and
Section \ref{evals} is a culmination of this fact and other results from the preceding sections, where we
construct certain invariants. We demonstrate the complexities inherent in adopting
the naive approach to evaluating the eigenvalues of these invariants by attempting to
express them in terms of the Casimir invariants of $gl(m|n+1)$ and $gl(m|n)$. 
We then follow up with the more
elegant approach using characteristic identities and vector shift operators to determine 
eigenvalue expressions. 
The main results are presented in Theorems \ref{TheoremC} and \ref{TheoremGAMMA} and Corollary
\ref{CorollaryDELTA}. Section \ref{evals} also provides some motivation, in the context of
{\em unitary} representations, for investigating these particular invariants by
considering reduced matrix elements and reduced Wigner coefficients.


\section{Preliminaries} \label{pp}

Throughout we adopt the graded index notation of Jarvis and Green \cite{JarGre1979}, where
Latin indices $1\leq i,j,k,\ell,\ldots\leq m$ are always assumed to correspond to ``even''
labels and Greek indices $1\leq \mu,\nu,\ldots\leq n$ are assumed ``odd''. We
associate with even and odd indices the parity factor
$$
(i) = 0,\ \ (\mu) = 1.
$$
This in fact corresponds to the standard $\mathbb{Z}_2$-gradation for the vector
representation.
Occasionally we find it convenient to introduce ungraded indices $1\leq p,q,r,s,\ldots\leq
m+n$ for the sake of uniformity of exposition.

The $gl(m|n)$ generators $E_{pq}$ ($1\leq p,q\leq m+n$) satisfy the graded
commutation relations
\beq
[E_{pq},E_{rs}] = \delta_{qr}E_{ps} - (-1)^{((p)+(q))((r)+(s))}\delta_{ps}E_{rq},
\label{eq1}
\eeq
where the graded commutator is given by
$$
[E_{pq},E_{rs}] = E_{pq}E_{rs} - (-1)^{((p)+(q))((r)+(s))}E_{rs}E_{pq}.
$$
Note in particular that this bracket satisfies graded antisymmetry, i.e.
$$
[E_{pq},E_{rs}] = -(-1)^{((p)+(q))((r)+(s))}[E_{rs},E_{pq}].
$$

A basis for the Cartan subalgebra $H$ of $gl(m|n)$ comprises the set of mutually commuting
generators $E_{pp}$ whose eigenvalues are employed to label the weights occuring in
the representations. Following Kac \cite{Kac1977}, we may expand a weight in terms
of the fundamental weights $\varepsilon_i$ ($1\leq i\leq m$) and $\delta_\mu$ ($1\leq \mu
\leq n$), which provides a convenient basis for $H^*$. Indeed we may expand a weight $\Lambda\in H^*$ as
$$
\Lambda = \sum_{i=1}^m\Lambda_i\varepsilon_i + \sum_{\mu=1}^n\Lambda_\mu\delta_\mu.
$$
With this convention, the root system is given by the set of even roots
\begin{align}
\pm(\varepsilon_i-\varepsilon_j), & \ \ 1\leq i<j\leq m,\nn\\
\pm(\delta_\mu - \delta_\nu), & \ \ 1\leq \mu<\nu\leq n,\nn
\end{align}
and the set of odd roots
\begin{align}
\pm(\varepsilon_i-\delta_\mu), \ \ 1\leq i\leq m,\ \ 1\leq \mu\leq n.
\end{align}
A system of simple roots is given by the distinguished set
$$
\left\{ \left. \varepsilon_i-\varepsilon_{i+1},\ \varepsilon_m-\delta_1,\
\delta_\mu-\delta_{\mu+1}\ \right| \ 1\leq i<m,\ 1\leq \mu<n \right\}.
$$
The sets of even and odd positive roots are then given, respectively, by
\begin{align}
\Phi_0^+ &= \{ \varepsilon_i - \varepsilon_j\ |\ 1\leq i<j\leq m\} \cup \{\delta_\mu -
\delta_\nu\ |\ 1\leq \mu<\nu\leq n\},\nn\\
\Phi_1^+ &= \{ \varepsilon_i - \delta_\mu\ |\ 1\leq i\leq m,\ 1\leq\mu\leq n\}. \nn
\end{align}
We set $\rho$ to be the graded half-sum of positive roots, i.e.
\begin{align}
\rho &= \frac12 \sum_{\alpha\in\Phi_0^+}\alpha - \frac12\sum_{\beta\in\Phi_1^+}\beta\nn\\
&= \frac12 \sum_{j=1}^m(m-n-2j+1)\varepsilon_j +
\frac12\sum_{\nu=1}^n(m+n-2\nu+1)\delta_\nu.
\label{rho}
\end{align}


Every finite dimensional irreducible representation of $gl(m|n)$ is a 
$\mathbb{Z}_2$-graded vector space
$$
V=V_0\oplus V_1,
$$
(so that $v\in V_j$ implies the grading $(v)=j$ for $j=0,1$)
and admits a highest weight vector, whose
weight $\Lambda$ uniquely characterises the representation. We denote the associated irreducible
highest weight module by $V(\Lambda)$ and the associated representation by
$\pi_\Lambda$. Relative to the $\mathbb{Z}_2$-grading, it is
assumed, unless stated otherwise, that the highest weight vector $v^\Lambda$ has an even
grading, i.e. $v^\Lambda\in V(\Lambda)_0$.
Components of the highest weight $\Lambda$ satisfy the lexicality conditions
$$
\Lambda_i - \Lambda_j\in \mathbb{Z}_+\ (1\leq i<j\leq m),\ \ \Lambda_\mu-\Lambda_\nu\in
\mathbb{Z}_+ \ (1\leq \mu<\nu\leq n), 
$$
but we note that $\Lambda_i+\Lambda_\mu$ may be any complex number.
As a simple example, the fundamental vector representation is denoted $V(\varepsilon_1)$ using this
notation. 


The fundamental vector representation $\pi_{\varepsilon_1}$ of $gl(m|n)$ is $(m+n)$ dimensional with a
basis $\{ \ket{p}\ |\ 1\leq p\leq m+n\}$ on which the generators $E_{pq}$ have the
following action:
$$
E_{pq}\ket{s} = \delta_{qs}\ket{p},
$$
so that
$$
\bra{r}E_{pq}\ket{s} = \delta_{qs}\bracket{r}{p} = \delta_{qs}\delta_{pr}
$$
or alternatively
$$
\pi_{\varepsilon_1}\left( E_{pq} \right)_{rs} = \delta_{qs}\delta_{pr}.
$$
This gives rise to a non-degenerate even invariant bilinear form on $gl(m|n)$ defined by
$$
(x,y) = \mbox{str}(\pi_{\varepsilon_1}(xy)) = \sum_{i=1}^m\pi_{\varepsilon_1}(xy)_{ii}
-\sum_{\mu=1}^n\pi_{\varepsilon_1}(xy)_{\mu\mu},
$$
where str denotes the supertrace given in \cite{Kac1978}.
In particular we have (sum over repeated indices)
\begin{eqnarray}
\left( E_{pq},E_{rs} \right) 
& = & (-1)^{(s)}\delta_{qr}\delta_{ps},
\label{eq2}
\end{eqnarray}
which leads to a bilinear form on the fundamental weights
$$
(\varepsilon_i,\varepsilon_j) = \delta_{ij},\ \ (\varepsilon_i,\delta_\mu)=0, \ \ (\delta_\mu,\delta_\nu) =
-\delta_{\mu\nu},
$$
which in turn induces a non-degenerate bilinear form on our weights $\Lambda$ given by
\begin{equation}
\left( \Lambda,\Lambda'\right) = \sum_{i=1}^m\Lambda_i\Lambda_i' -
\sum_{\mu=1}^n\Lambda_\mu\Lambda_\mu'.
\label{form}
\end{equation}

Note that the left dual basis of $\{E_{pq}\}$ under the form (\ref{eq2}) is
given by $\{(-1)^{(q)}E_{qp}\}$, i.e.
$$
\left( (-1)^{(r)}E_{rs},E_{pq} \right) =\delta_{qr}\delta_{ps}. 
$$
It follows that the second order universal Casimir invariant is given by
\beq
I_2 = (-1)^{(q)}E_{pq}E_{qp},
\label{eq3}
\eeq
(summation over repeated indices assumed) which is a well-defined element of the universal
enveloping algebra $U$ of $gl(m|n)$. Indeed, it may be verified directly
that $I_2$ is central, i.e.
\begin{eqnarray*}
I_2 E_{rs} - E_{rs}I_2 
& = & 0.
\end{eqnarray*}
We can expand $I_2$ as
$$
I_2 = \sum_{i,j}E_{ji}E_{ij} + \sum_{i,\mu}E_{\mu i}E_{i\mu} 
- \sum_{i,\mu}E_{i\mu}E_{\mu i} - \sum_{\mu,\nu}E_{\mu\nu}E_{\nu\mu}.
$$
Since $I_2$ is central, it must take a constant value on any (standard cyclic) highest weight module. 
For a module with highest weight vector $v$ corresponding to highest weight $\Lambda$ we have 
\begin{eqnarray*}
\pi_\Lambda(I_2) v & = & \sum_{j\leq i}\pi_\Lambda(E_{ji}E_{ij})v - \sum_{i,\mu}\pi_\Lambda(E_{i\mu}E_{\mu i})v -
\sum_{\nu\leq \mu}\pi_\Lambda(E_{\nu\mu}E_{\mu\nu})v\\
& = & \sum_{i=1}^m\Lambda_i^2v + \sum_{i=1}^m(m-(2i-1))\Lambda_iv - n\sum_{i=1}^m\Lambda_iv
\\
&&\quad - m\sum_{\mu=1}^n\Lambda_\mu v - \sum_{\mu=1}^n(n-(2\mu-1))\Lambda_\mu v -
\sum_{\mu=1}^n\Lambda_\mu^2v
\end{eqnarray*}
so that the eigenvalue of $I_2$, denoted $\chi_\Lambda(I_2)$ is given by
$$
\chi_\Lambda(I_2) = \sum_{i=1}^m \Lambda_i(\Lambda_i+m-n-2i+1) 
- \sum_{\mu=1}^n\Lambda_\mu(\Lambda_\mu+m+n-2\mu+1).
$$
Making use of equations (\ref{rho}) and (\ref{form}),
this may be conveniently expressed
\begin{align}
\chi_\Lambda(I_2) = (\Lambda,\Lambda+2\rho). \label{caseig}
\end{align}


\section{Characteristic identities}\label{ci}

If $\pi_\theta$ denotes a finite dimensional irreducible representation of $gl(m|n)$ with
highest weight $\theta$, we may construct the tensor matrix $A^\theta$ with algebraic entries
$$
A^\theta_{\alpha \beta} 
= - \sum_{p,q} (-1)^{(q)+((p) + (q)) (\beta)} \pi_\theta (E_{pq})_{\alpha \beta} E_{qp}
$$
where $\{ {e_\alpha } \}$ is a fixed homogeneous basis for the $gl(m|n)$ module
$V(\theta)$. Acting on a finite dimensional irreducible $gl(m|n)$ module $V(\Lambda)$ the
matrix $A^\theta$ may be expressed in the invariant form
$$
A^\theta = - \frac{1}{2} \left[ (\pi_\theta \otimes \pi_\Lambda)(\Delta(I_2)) -
\pi_\theta (I_2) \otimes I - I \otimes \pi_\Lambda
(I_2)\right],
$$
where $\Delta:U\rightarrow U\otimes U$ is the usual coproduct on the universal enveloping algebra $U$, 
and $I$ denotes the identity matrix on $V(\Lambda),$ $V(\theta)$ respectively.
If $\theta_1,\theta_2,\ldots,\theta_k$ denote the {\em distinct} weights occurring in $V(\theta)$, it
follows from previous work of Gould \cite{Gould1987} that the tensor matrix $A^\theta$ satisfies the following
polynomial identity on $V(\Lambda)$:
$$
\prod_{i=1}^k (A^\theta - \alpha_i) = 0
$$
where 
\begin{align}
\alpha_i &= -\frac{1}{2} [ \chi_{\Lambda+\theta_i} (I_2) - \chi_\theta (I_2) -
\chi_\Lambda (I_2)]\nn\\
&= \frac{1}{2} (\theta,\theta+2\rho) - \frac{1}{2} (\theta_i,\theta_i + 2(\Lambda + \rho)).
\label{eq:ROOTS}
\end{align}

We are concerned here with the vector representation $\pi_{\varepsilon_1}$ (with
$\theta=\varepsilon_1$) 
in which case the tensor matrix $A^{\varepsilon_1}$ is given by
\begin{align}
A^{\varepsilon_1}_{p q} &= - \sum_{r,s} (-1)^{(s)+((r) + (s))(q)}
\pi_{\varepsilon_1}(E_{rs})_{p q}
 E_{sr}\nn\\
&= - (-1)^{(p)(q)}E_{q p},\nn
\end{align}
where indices $1 \leq p,q \leq m+n$ are assumed ungraded. 
We thus obtain what we call the $gl(m|n)$ adjoint matrix:
\begin{align}
{\cal \bar{A}}_p^{\ q} \equiv A^{\varepsilon_1}_{p q}= - (-1)^{(p)(q)}E_{q p}.\label{adjoint}
\end{align}

The weights in the representation $\pi_{\varepsilon_1}$ are of the form $\varepsilon_i$ $(1 \leq i \leq m)$,
$\delta_\mu$ $(1 \leq \mu \leq n)$ from which it follows that the adjoint matrix
${\cal \bar{A}}$
satisfies the characteristic identity
\begin{align}
\prod_{i=1}^m ({\cal \bar{A}} - \bar{\alpha}_i) \prod^n_{\mu = 1} ({\cal \bar{A}} - \bar{\alpha}_\mu) =
0
\label{notes6}
\end{align}
when acting on an irreducible $gl(m|n)$ module $V(\Lambda)$, where the adjoint roots
$\bar{\alpha}_i, \bar{\alpha}_\mu$ are given, in accordance with equation~(\ref{eq:ROOTS}), by
\begin{align}
\bar{\alpha}_r = -\frac{1}{2} [ \chi_{\Lambda+\varepsilon_r}(I_2) -\chi(I_2) -\chi_{\Lambda}(I_2)]
\nn
\end{align}
where $\chi(I_2) = m-n$ is the eigenvalue of $I_2$ on the vector representation. 
Using equation (\ref{rho}) we thus obtain
\begin{align}
\bar{\alpha}_i &= -\frac{1}{2} [\chi_{\Lambda+\varepsilon_i}(I_2) + n - m -
\chi_\Lambda(I_2)]\nn\\
&=-\frac{1}{2} [ (\varepsilon_i,\varepsilon_i)  + 2(\varepsilon_i,\Lambda + \rho) + n -m]\nn\\
& = i - 1 -\Lambda_i.\label{eq:adjCHAR_ROOTSa}
\end{align}
Similarly for the odd adjoint roots we obtain
\begin{align}
\bar{\alpha}_\mu &= -\frac{1}{2} [ \chi_{\Lambda+\delta_\mu}(I_2) + n - m -
\chi_\Lambda(I_2)]\nn\\
&= \Lambda_\mu + m + 1 - \mu.\label{eq:adjCHAR_ROOTSb}
\end{align}

To construct the $gl(m|n)$ vector matrix we take $\tilde{\pi}$ to be the triple dual of
the vector representation (viz. $\tilde{\pi} = \pi_{\varepsilon_1}^{***}$ ) defined by
\begin{align}
\tilde{\pi} (E_{pq})_{rs} = -(-1)^{(p)((p) + (q))} \delta_{qr} \delta_{ps}.\nn
\end{align}
Our previous construction for the matrix $A^\theta$ with $\pi_{\theta}$ replaced by
$\tilde{\pi}$ yields the $gl(m|n)$ vector matrix 
\begin{align}
	{\cal A}^{p}_{\ q} &= - \sum_{r,s} (-1)^{(s)+((r) + (s))(q)} \tilde{\pi}
(E_{rs})_{p q} E_{sr}\nn\\
	&= (-1)^{(p)}E_{p q}.\label{equdefx}
\end{align}
The weights occurring in $\tilde {\pi}$ are the $-\varepsilon_i (1 \leq i \leq m),
-\delta_\mu (1\leq \mu \leq n)$ from which it follows that acting on the irreducible
$gl(m|n)$ module $V(\Lambda)$, the matrix ${\cal A}$ satisfies the characteristic identity
\begin{align}
 \label{eq:POLY_IDENT}
	\prod^m_{i=1} ({\cal A} - \alpha_i) \prod^n_{\mu=1} ({\cal A} - \alpha_\mu) = 0 
\end{align}
where our characteristic roots are given by
\begin{align}
\alpha_i &= -\frac{1}{2} [ \chi_{\Lambda-\varepsilon_i}(I_2) + n - m -
\chi_\Lambda (I_2)]\nn\\
&= \Lambda_i + m - n - i,\ \ 1\leq i\leq m,
\label{eq:CHAR_ROOTSa}
\\
\alpha_\mu &= -\frac{1}{2} [ \chi_{\Lambda-\delta_\mu}(I_2) + n - m - \chi_\Lambda
(I_2)]\nn\\
&= \mu-\Lambda_\mu - n,\ \ 1\leq \mu \leq n.
\label{eq:CHAR_ROOTSb}
\end{align}
We remark that in the above we used the fact that the eigenvalue of $I_2$ in the
representation $\tilde{\pi}$ is given by
\begin{align}
	\tilde{\chi}(I_2) = m-n, \nn
\end{align}
which is the same as the eigenvalue in the vector representation. 

{\bf Remarks:}
\begin{enumerate}
\item Note that
\begin{align}
\alpha_i - \alpha_\mu + 1 = (\Lambda+\rho,\varepsilon_i-\delta_\mu) = -(\bar{\alpha}_i
-\bar{\alpha}_\mu + 1).\nn
\end{align}
For {\em atypical} irreducible representations \cite{Kac1978}, we have 
\begin{align}
(\Lambda+\rho,\varepsilon_i-\delta_\mu)=0, \label{oddroot}
\end{align}
so that the equation $\alpha_\mu = \alpha_i+1$ (or $\bar{\alpha}_\mu = \bar{\alpha}_i+1$)
for some $i$ and $\mu$ serves as an atypicality condition in terms of the characteristic
roots. Equation (\ref{oddroot}) also implies that
\begin{align}
\alpha_i-\alpha_\mu &= (\Lambda-\varepsilon_i+\rho,\varepsilon_i-\delta_\mu),\nn\\
\bar{\alpha}_i - \bar{\alpha}_\mu &= -(\Lambda+\varepsilon_i+\rho,\varepsilon_i-\delta_\mu).\nn
\end{align}
\item In the case we take $\pi_\theta= \pi_{\varepsilon_1}^*$ (the dual of the vector representation) we
obtain a new matrix, referred to as the double adjoint:
\begin{align}
{\cal \bar{\bar{A}}}_{r q} = (-1)^{(q)}E_{r q} = (-1)^{(r) + (q)}
{\cal A}_{r q}.\nn
\end{align}
This is the matrix appearing in the work of Jarvis and Green \cite{JarGre1979}. If $p(x)$
is any polynomial we have (e.g. by induction on the degree of $p(x)$)
\begin{align}
p\left({\cal \bar{\bar{A}}}\right)_{r q} = (-1)^{(r) + (q)} p({\cal A})_{r q}\nn
\end{align}
so that ${\cal \bar{\bar{A}}}$ and ${\cal A}$ satisfy the same characteristic identity.
Equations~(\ref{eq:CHAR_ROOTSa}) and (\ref{eq:CHAR_ROOTSb}) agree with the roots obtained
by Jarvis and Green using different methods, except that equation~(\ref{eq:CHAR_ROOTSb}) 
correctly replaces $-\mu$ with $\mu$ in the formula of Jarvis and Green.
\item We also have a triple adjoint matrix defined by 
\begin{align}
{\cal \bar{\bar{\bar{A}}}}_p^{\ q} = (-1)^{(p) + (q)} {\cal \bar{A}}_p^{\ q}\nn
\end{align}
which satisfies the same characteristic identity as ${\cal \bar{A}}$. Note that the matrix
${\cal \bar{\bar{A}}}$ (resp. ${\cal \bar{\bar{\bar{A}}}}$) is simply related to ${\cal A}$ (resp. ${\cal \bar{A}}$)
by the $gl(m|n)$ grading automorphism.
\item Polynomials in ${\cal A}$ and ${\cal \bar{A}}$ are defined recursively according to
\begin{align}
({\cal A}^{k+1})^p_{\ q} &= \sum_r({\cal A}^k)^p_{\ r}\  {\cal A}^r_{\ q} = \sum_r
{\cal A}^p_{\ r}\ ({\cal A}^k)^r_{\ q}\nn\\
({\cal \bar{A}}^{k+1})_p^{\ q} &= \sum_r({\cal \bar{A}}^k)_p^{\ \
r}\ 
{\cal \bar{A}}_r^{\ q}
= \sum_r {\cal \bar{A}}_p^{\ r} \ ({\cal \bar{A}}^k)_r^{\ q}.\nn
\end{align}
\end{enumerate}


\section{Tensor operators}  \label{to}

Let $V(\Lambda)$ be a finite dimensional irreducible $gl(m|n)$ module with highest weight
$\Lambda$ and let
\begin{align}
T: V(\Lambda) \otimes V \longrightarrow W\nn
\end{align}
be a (surjective) intertwining operator of degree $(\tau)$ where $V$ and $W$  are $gl(m|n)$ modules:
\begin{align}
\pi_W(E_{p q})\ T = (-1)^{((p) + (q))(\tau)}\ T\  
(\pi_\Lambda \otimes \pi)\Delta(E_{p q})\nn
\end{align}
 where $\pi$ (resp. $\pi_W$) is the representation of $gl(m|n)$ afforded by $V$ (resp.
$W$). Now let $\{ e_\alpha\}$ be a homogeneous basis for $V(\Lambda)$. Then we may
define a collection of operators $\{T_\alpha\}$, called a \textit{tensor operator},
operating on $V$ according to
\begin{align}
T_\alpha\ v = T (e_\alpha \otimes v)~~~,\forall v \in V\nn
\end{align}
with $T$ as above.
We have, for arbitrary $v \in V$, and homogeneous $x \in gl(m|n)$
\begin{align}
x\ T_\alpha\ v &\equiv \pi_W(x)\ T\ (e_\alpha \otimes v)\nn\\
&= (-1)^{(x)(\tau)} T (\pi_\Lambda (x) \otimes \pi(I) + \pi_\Lambda(I) \otimes \pi(x) )
(e_\alpha \otimes v)\nn\\
&= (-1)^{(x)(\tau)} T \left( \pi_\Lambda (x)_{\beta \alpha} e_\beta \otimes v +
(-1)^{(\alpha)(x)} e_\alpha \otimes \pi(x)v \right)\nn \\
&\equiv \left((-1)^{(x)(\tau)} \pi_\Lambda (x)_{\beta \alpha} T_\beta + (-1)^{(x)((\tau) +
(\alpha))} T_\alpha\ x \right) v.\nn
\end{align}
\begin{enumerate}
\item[] {\bf Note:} We have utilised the summation convention over repeated indices. We will adopt this
convention throughout the paper.
\end{enumerate}
Thus, by abstraction, we define an irreducible tensor operator of rank $\Lambda$ and
degree $(\tau)$ as a collection of components $\left\{ T_\alpha \right\}$ transforming
according to
\begin{align}
[x,T_\alpha] = (-1) ^{(x)(\tau)} \pi_\Lambda (x)_{\beta \alpha}\ T_\beta
\label{eq:TENSOR_OPERATOR}
\end{align}
where $(x)$ is the degree of $x \in gl(m|n)$ and the graded bracket on the left hand side
is given by
\begin{align}
[x,T_\alpha] = x\ T_\alpha - (-1)^{(x)((\alpha) + (\tau))} T_\alpha\ x.\nn
\end{align}
In the special case where $\pi_\Lambda = \pi_{\varepsilon_1}$ is the vector representation we obtain the
transformation law of vector operators $\psi^r$ $(1 \leq r \leq m+n)$:
\begin{align}
[E_{pq},\psi^r] &= (-1)^{(\psi)((p) + (q))} \pi_{\varepsilon_1}(E_{pq})_{sr} \psi^s\nn\\
&= (-1)^{(\psi)((p) + (q))} \delta^r_q \psi^p\nn
\end{align}
If $(\psi) = 0$ (resp. $1$) we call $\psi$ an even (resp. odd) vector operator:
the case $(\psi) = 0$ corresponds to the definition of vector operator given by Jarvis and
Green \cite{JarGre1979}. In the case that $\pi_\Lambda = \pi_{\varepsilon_1}^*$ is the dual of the vector representation we
obtain the transformation law of contragredient vector operators $\phi_r$ $(1\leq r\leq
m+n)$:
\begin{align}
[E_{pq},\phi_r] = (-1)^{(\phi)((p)+(q))} \pi_{\varepsilon_1}^* (E_{pq})_{sr} \phi_s\nn
\end{align}
where
\begin{align}
\pi_{\varepsilon_1}^*(E_{pq})_{sr} &\stackrel{\mbox{def.}}{=} -(-1)^{(s)((p) + (q))}
\pi_{\varepsilon_1}(E_{pq})_{rs}\nn\\
&= - (-1)^{(q)((p) + (q))} \delta_{pr} \delta_{qs}\nn
\end{align}
\begin{align}	
\label{eq:CONTRA_TRANSFORM}
\Rightarrow [E_{pq},\phi_r] = -(-1)^{((p) + (q))((\phi) + (q))} \delta_{pr} \phi_q
\end{align}
If $(\phi) = 0$ (resp. $1$) we say that $\phi$ is an even (resp. odd) homogeneous contragredient vector
operator. Our main concern here, is with $gl(m|n)$ vector and
contragredient vector operators, whose transformation laws are given above. We should
remark that if we take $\pi_\Lambda$ in equation~(\ref{eq:TENSOR_OPERATOR}) to be one of the
tensor representations, then appropriate transformation laws for higher order tensor
operators can be given.

{\bf Remarks:}
\begin{enumerate}
\item
If $\psi^r$ is a vector operator then
\begin{align}
\tilde{\psi}^r = (-1)^{(r)} \psi^r\nn
\end{align}
also constitutes a tensor operator whose components $\tilde{\psi}^r$ transform according
to the double dual $\pi_{\varepsilon_1}^{**}$ of the vector representation
$\pi_{\varepsilon_1}$. Similarly if $\phi_r$
transforms as in equation~(\ref{eq:CONTRA_TRANSFORM}) then
\begin{align}
\tilde{\phi}_r = (-1)^{(r)} \phi_r\nn
\end{align}
constitutes a tensor transforming as the triple dual $\pi_{\varepsilon_1}^{***}$ of the vector
representation $\pi_{\varepsilon_1}$.
\item
An {\em odd} vector operator is equivalent to an even vector operator but with a reversal
of the $\mathbb{Z}_2$-grading in the vector representation.
\end{enumerate}

\section{Vector operator shift components} \label{vosc}

Since
$\psi^p$ transforms as the vector representation $\pi_{\varepsilon_1}$ it follows that the $gl(m|n)$
adjoint matrix (\ref{adjoint}) acts naturally on the {\em right} of $\psi$:
\begin{align}
(\psi {\cal \bar{A}})^p = \psi^q { \cal\bar{A}}_q^{\ p}.
\end{align}
We may then proceed to resolve $\psi$ into its shift components via the use of projections
\cite{Green1971,BraGre1971}:
\begin{align}
\psi^p = \sum^m_{i=1} \psi[i]^p + \sum^n_{\mu = 1} \psi[\mu]^p \nn
\end{align}
where the shift components $\psi[r]$ are constructed as follows (single ungraded index
notation in use):
\begin{align}
\label{eq:SHIFT_COMPONENTS_PSI}
\psi[r]^p = \psi^q \bar{P} [r]_q^{\ p} ~~~~1\leq r \leq m+n
\end{align}
where
\begin{align}
\bar{P}[r] = \prod^{m+n}_{k \neq r} \left( \frac{{\cal \bar{A}} - \bar{\alpha}_k}{\bar{\alpha}_r
- \bar{\alpha}_k} \right)\nn
\end{align}
is the appropriate projection operator constructed using the characteristic identity~(\ref{notes6}). 

Before proceeding we show that the shift
components~(\ref{eq:SHIFT_COMPONENTS_PSI}) of a vector operator indeed constitute a vector
operator. Using a simple induction argument, since 
$\bar{P}[r]$ 
is a polynomial in ${\cal \bar{A}}$, it suffices to show that if $\psi^p$ is a homogeneous vector
operator, then so too is
$$
(\psi {\cal \bar{A}})^p = \psi^q {\cal \bar{A}}_q^{\ p} =
-(-1)^{(p)(q)}\psi^q E_{pq}.
$$ 
We have
\begin{align}
[E_{pq},(\psi {\cal \bar{A}})^r] &=
-(-1)^{(r)(s)}[E_{pq},\psi^s E_{rs}]\nn\\
&= -(-1)^{(r)(s)} \left( [E_{pq},\psi^s]E_{rs} +
(-1)^{((p)+(q))((s)+(\psi))}
\psi^s[E_{pq},E_{rs}]\right)\nn\\
&= 
-(-1)^{(r)(s)} 
\left( \delta^s_q \psi^p (-1)^{((p)+(q))(\psi)}E_{rs} +
\delta^r_q
(-1)^{((p)+(q))((s)+(\psi))}
\psi^s E_{ps}\right.\nn\\
& \qquad 
\left.- \delta^p_s(-1)^{((p)+(q))((r)+(\psi))}
\psi^s E_{rq} \right)
\nn\\
&=
-(-1)^{((p)+(q))(\psi)+(p)(s)}\delta^r_q\psi^s
E_{ps}\nn\\
&= (-1)^{((p)+(q))(\psi)}
\delta_q^r\psi^s{\cal \bar{A}}_s^{\ p}\nn\\
&=  (-1)^{((p)+(q))(\psi)}
\delta_q^r(\psi {\cal \bar{A}})^p.\nn
\end{align}
Thus $\psi{\cal \bar{A}}$ is also a homogeneous vector operator of degree $(\psi)$.
Alternatively note that $\bar{P}[r]$ determines an intertwining operator and
hence so too does $\psi[r]=\psi \bar{P}[r]$. It follows that the components of $\psi[r]$
must also determine a vector operator of the same degree.

{\bf Remark:} 
\begin{enumerate}
\item[] 
At this point we highlight the fact that for certain irreducible representations, the characteristic
roots may coincide (consider $\bar{\alpha}_r=\bar{\alpha}_k$ in the above formula). This
is related to the occurrence of atypical irreducible representations in the tensor product
of $V({\varepsilon_1})\otimes V(\Lambda)$ or $V({\varepsilon_1})^*\otimes V(\Lambda)$. The
set of $\Lambda$ for which this happens, however, is closed in the Zariski topology
\cite{Humphreys} on $H^*$. 
Therefore the roots of the characteristic identity are distinct on an open and hence dense subset of
$H^*$. Hence without loss of generality, we will make the assumption that the roots are
distinct throughout the remainder of the paper unless otherwise indicated. In fact it can
be shown that under this assumption, the tensor products $V({\varepsilon_1})\otimes V(\Lambda)$ 
and $V({\varepsilon_1})^*\otimes V(\Lambda)$ are completely reducible (see Appendix
B for details). Furthermore, it
is worth remarking that the invariants to be evaluated in this paper determine (rational)
polynomial functions which are continuous in the Zariski topology. Note however, care
needs to be taken when applying our formulae, by first cancelling terms in numerators and
denominators where appropriate. 
\end{enumerate}

From the previous work of Gould \cite{Gould1981}, the above shift components (\ref{eq:SHIFT_COMPONENTS_PSI}) 
effect the following shifts in the representation labels $\Lambda$:
\begin{align}
\psi[i]:&~\Lambda_j \rightarrow \Lambda_j + \delta_{ij} ~~~(1 \leq i,j \leq m),\nn\\
\psi[\mu]:&~\Lambda_\nu \rightarrow \Lambda_\nu + \delta_{\nu \mu} ~~~(1 \leq \mu,\nu \leq
n),\nn
\end{align}
the remaining labels remaining unchanged. 

In a similar way, if $\phi_p$ is a contragredient vector operator then the matrix
${\cal \bar{\bar{A}}}$ acts naturally on the right of $\phi$:
\begin{align}
(\phi {\cal \bar{\bar{A}}})_p = \phi_q ({\cal \bar{\bar{A}}})^q_{\ p} =
(-1)^{(p)+(q)}\phi_q
{\cal A}^q_{\ p}.\nn
\end{align}
Thus we obtain the resolution
\begin{align}
\label{eq:PHI_RESOLUTION}
\phi_p = \sum^m_{i=1} \phi[i]_p + \sum^n_{\mu = 1} \phi [\mu]_p
\end{align}
where
\begin{align}
\label{eq:SHIFT_COMPONENTS_PHI}
\phi[r]_p = (-1)^{(p)+(q)}\phi_q P[r]^q_{\ p} 
\end{align}
where our $gl(m|n)$ vector projectors are given by
\begin{align}
P[r] = \prod^{m+n}_{k \neq r} \left( \frac{{\cal A} - \alpha_k}{\alpha_r - \alpha_k} \right)\nn
\end{align}

In this case the shift components~(\ref{eq:SHIFT_COMPONENTS_PHI}) effect the following
shifts on the representation labels:
\begin{align}
&\phi[i]:~\Lambda_j \rightarrow \Lambda_j - \delta_{ij} ~~~(1 \leq i,j \leq m)\nn\\
&\phi[\mu]:~\Lambda_\nu \rightarrow \Lambda_\nu - \delta_{\mu \nu} ~~~(1 \leq \mu,\nu \leq
n),\nn
\end{align}
the other labels remaining unchanged.

We remark that the shift components~(\ref{eq:PHI_RESOLUTION}) of a
contragredient vector $\phi_r$ indeed constitute a contragredient vector, in a
similar way to the case of vector operators.

The results above all hold regardless of whether our vector operators are even or odd.
However the matrices which act on the {\em left} of vectors and contragredient vectors
will depend explicitly on their degree (odd or even). We shall be primarily concerned with
{\em odd} vector and contragredient vector operators in this paper so we shall
concentrate on them in the following (although an analogous formalism can be set up for
the even case $(\tau) = 0$).

It turns out that odd vector (and contragredient vector) operators appear naturally in
discussing the Lie superalgebra embedding $gl(m|n+1)\supset gl(m|n)$. 
Throughout the remainder of the paper we assume, unless otherwise stated, that $\psi^r$
(resp. $\phi_r$) denotes a $gl(m|n)$ odd vector (resp. odd contragredient vector)
operator. That is, we assume the transformation laws
\begin{align}
[E_{pq},\psi^r] &= (-1)^{(p) + (q)} \delta^r_q \psi^p,\nn\\
[E_{pq},\phi_r] &= -(-1)^{(p)((p) + (q))} \delta_{pr} \phi_q.\nn
\end{align}
We note that the graded brackets on the left hand side are given by
\begin{align}
[E_{pq},\psi^r] &= E_{pq} \psi^r -(-1)^{((p) + (q))((r) + 1)} \psi^r E_{pq} \nn\\
&= -(-1)^{((p) + (q))((r) + 1)}[\psi^r,E_{pq}]\nn
\end{align}
and similarly
\begin{align}
[E_{pq},\phi_r] &=  -(-1)^{((p) + (q))((r) + 1)}[\phi_r,E_{pq}],\nn
\end{align}
since we are assuming that $(\psi)=1=(\phi)$.

Since the matrix ${\cal \bar{A}}$ acts naturally on the right of $\psi^p$ we have
\begin{align}
(\psi {\cal \bar{A}})^p &= \psi^q {\cal \bar{A}}_q^{\ p}\nn\\
&= -(-1)^{(p)(q)} \psi^q E_{p q}\nn\\
&= -(-1)^{(p)(q)} \left( [\psi^q,E_{p q}] + (-1)^{[(p)
+ (q)][(q) + 1]} E_{p q} \psi^q \right)\nn\\
&= (-1)^{(p)} \left( [E_{p q}, \psi^q ] - E_{p q} \psi^q
\right)\nn\\
&= (-1)^{(q)} \delta^q_q \psi^p - (-1)^{(p)}E_{p q}
\psi^q\nn\\
&= (m - n - {\cal A})^p_{\ q} \psi^{q}\nn
\end{align}
It follows that the $gl(m|n)$ matrix ${\cal A}$ acts naturally on the left of odd vector
operators (while the double adjoint ${\cal \bar{\bar{A}}}$ acts naturally on the left of even
vector operators).

Similarly, for contragredient vectors $\phi_p$ we have 
\begin{align}
(\phi {\cal \bar{\bar{A}}})_p &= \phi_q {\cal \bar{\bar{A}}}^q_{\ p}\nn\\
&= (-1)^{(p)} \phi_q E_{q p}\nn\\
&= (-1)^{(p)} \left( [\phi_q,E_{q p} ] + (-1)^{((p) +
(q))((q) + 1)} E_{q p} \phi_q \right)\nn\\
&= (-1)^{(p)(q)} \left( E_{q p } \phi_q - [E_{q p},
\phi_q] \right)\nn\\
&= (-1)^{(p)(q)} E_{q p} \phi_q  + (-1)^{(q)} \delta^q_q
\phi_p \nn\\
&= (m - n - {\cal \bar{A}})_p^{\ q} \phi_{q}.\nn
\end{align}
It thus follows that the matrix ${\cal \bar{A}}$ acts naturally on the left of odd contragredient
vectors.

Thus we may project out the shift components of $\psi$ and $\phi$ from the left according
to
\begin{align}
\psi[r]^p = P[r]^p_{\ q} \psi^q, ~~~ \phi[r]_p = \bar{P}
[r]_p^{\ q} \phi_q\label{ShiftLeft}
\end{align}
where $P[r],$ $\bar{P}[r]$ are the projection operators previously constructed in terms of
the matrices ${\cal A}$, ${\cal \bar{A}}$ respectively.


\section{Branching conditions for $gl(m|n+1)\supset gl(m|n)$} \label{br}

We now seek to determine necessary conditions for the branching rule $gl(m|n+1)\downarrow
gl(m|n)$. They turn out to be similar in appearance to the betweenness
conditions of \cite{StoiVan2010} for example, but here we give a detailed proof of the
necessary condition (but not sufficient) in a more general setting.

We first establish some notation. We set
\begin{align*}
\hat{L} = gl(m|n+1) = \hat{L}_- \oplus \hat{L}_0 \oplus \hat{L}_+
\end{align*}
where
\begin{align*}
\hat{L}_0 = gl(m) \oplus gl(n+1),
\end{align*}
and similarly
\begin{align*}
L = gl(m|n) \oplus gl(1) = L_- \oplus L_0 \oplus L_+
\end{align*}
where
\begin{align*}
L_0 = gl(m) \oplus gl(n) \oplus gl(1).
\end{align*}
We also introduce the $L_0$-modules
\begin{align*}
K_+ = \hbox{span}~\{ E_{i,m+n+1} \}^m_{i=1},~K_- = \hbox{span}~\{ E_{m+n+1,i} \}^m_{i=1}
\end{align*}
so that
\begin{align*}
\hat{L}_{\pm} = L_{\pm} \oplus K_{\pm}.
\end{align*}


We now recall some facts about the representation theory of the algebra $\hat{L}$. First, every
finite-dimensional irreducible $\hat{L}$-module $V(\tilde{\Lambda})$ with highest weight
\begin{align*}
\tilde{\Lambda} = \sum^m_{i=1} \tilde{\Lambda}_i \epsilon_i + \sum^{n+1}_{\mu = 1}
\tilde{\Lambda}_\mu \delta_\mu
\end{align*}
admits a $\mathbb{Z}$-gradation
\begin{align}
 V(\tilde{\Lambda}) = \bigoplus^0_{i=-d} V_i (\tilde{\Lambda})
\label{equ3}
\end{align}
in which case we say that $V(\tilde{\Lambda})$ admits $d+1$ levels.
Here, $V_0(\tilde{\Lambda})$ is called the maximal $\mathbb{Z}$-graded component which
constitutes an irreducible $\hat{L}_0$-module of the same highest weight. We also observe that the
decomposition (\ref{equ3}) is in fact an $\hat{L}_0$-module decomposition.

Next we note \cite{Kac1978} that any such irreducible $\hat{L}_0$-module $V_0(\tilde{\Lambda})$ admits
an invariant inner product $\langle~,~\rangle$ which extends in a unique way to an invariant
non-degenerate sesquilinear form $\langle~,~\rangle$ on all of $V(\tilde{\Lambda})$ which satisfies
the symmetry
\begin{align*}
\langle v,w\rangle = \overline{\langle w,v\rangle} ~,~\forall v,w \in V(\tilde{\Lambda})
\end{align*}
and the invariance condition given by
\begin{align}
\langle av,w\rangle = \langle v,a^\dagger w\rangle~,~\forall a \in \hat{L}
\label{equ4}
\end{align}
where $\dagger$ is the conjugation operation defined on $\hat{L}$ by
\begin{align*}
(E_{pq})^\dagger = E_{qp}.
\end{align*}
Such an invariant sesquilinear form has all the properties of an inner product
except it is not necessarily positive definite. When it is, we call $V(\tilde{\Lambda})$
{\em unitary} of type 1.

{\bf Remarks:}
\begin{enumerate}
\item We may define matrix elements, Wigner coefficients etc. even for non-unitary
irreps in the usual way -- except we work with a non-degenerate sesquilinear form
(\ref{equ4}) rather that an inner product.
\item We also have another conjugation operation on $\hat{L}$ defined by 
\begin{align}
(E_{pq})^\dagger = (-1)^{(p) + (q)}E_{qp}.
\label{equ5b}
\end{align}
Then $V(\tilde{\Lambda})$ also admits a unique invariant non-degenerate sesquilinear form
(\ref{equ4}) with respect to the conjugation operation (\ref{equ5b}). When this form is
positive definite we call
$V(\tilde{\Lambda})$ {\em unitary} of type 2. It can be shown \cite{ZhaGou1990} that the two types of
unitary irreps are related by duality.
\end{enumerate}

We observe a number of properties of the form $\langle,\rangle$ of equation (\ref{equ4}). First the
decomposition (\ref{equ3}) is orthogonal with respect to the form. Next $V(\tilde{\Lambda})$ decomposes into
a direct sum of irreducible $L_0$ and $\hat{L}_0$ modules. Two irreducible $L_0$
(respectively $\hat{L}_0$) modules with different highest weights are necessarily orthogonal under the
form.

Before we present the main result of this section, we 
first note from the PBW theorem that 
\begin{align}
V(\tilde{\Lambda}) &= U(\hat{L}_-)V_0(\tilde{\Lambda})\nn\\
&= U(L_-)U(K_-)V_0(\tilde{\Lambda})\nn\\
&= U(L_-)W(\tilde{\Lambda}) \label{equ6new}
\end{align}
where 
$$
W(\tilde{\Lambda}) = U(K_-)V_0(\tilde{\Lambda}).
$$ 
This leads to the following useful Lemma.
\begin{lemma} \label{maxwtlemma} 
Let $0 \neq v_+ \in V(\tilde{\Lambda})$ be an $L$-maximal weight vector. Then we have
\begin{align*}
\langle v_+,W(\tilde{\Lambda}) \rangle \neq (0).
\end{align*}
\end{lemma}
\proof{
Otherwise we would have
\begin{align*}
(0) &= \langle U(L_+) v_+, W(\tilde{\Lambda}) \rangle\\
&\stackrel{(\ref{equ4})}{=} \langle v_+, U(L_-) W(\tilde{\Lambda})\rangle\\
&\stackrel{(\ref{equ6new})}{=}\langle v_+,V(\tilde{\Lambda})\rangle ~~\Rightarrow ~~ v_+ = 0,
\end{align*}
a contradiction.
} 

We are now in a position to prove our main result:

\begin{thm} \label{TheoremBR} 
Let $v_+ \in V(\tilde{\Lambda})$ be an $L$-maximal weight vector of weight
\begin{align*}
\Lambda = \sum^m_{i=1} \Lambda_i \epsilon_i + \sum^n_{\mu = 1} \Lambda_\mu \delta_\mu.
\end{align*}
Then:
\begin{enumerate}
\item[(i)] The components of $\Lambda$ must satisfy the betweenness conditions
\begin{align}
&\tilde{\Lambda}_\mu \geq \Lambda_{\mu} \geq \tilde{\Lambda}_{\mu+1}~~~,~1 \leq \mu
\leq n \nn\\
&\tilde{\Lambda}_i \geq \Lambda_i \geq \tilde{\Lambda}_i - 1 ~~~,~1 \leq i \leq m.
\label{equ6}
\end{align}
\item[(ii)] $v_+$ is the unique (up to scalar multiples) $L$-maximal weight vector in
$V(\tilde{\Lambda})$ of weight $\Lambda$.
\end{enumerate}
\end{thm}

{\bf Proof:}
To prove the theorem we 
note that $W(\tilde{\Lambda})$ of equation (\ref{equ6new}) gives rise to a $L_0$-module and decomposes
into a direct sum of irreducible $L_0$ submodules
\begin{align}
W(\tilde{\Lambda}) = \bigoplus_\Lambda \hat{V}_0(\Lambda)
\label{equ7}
\end{align}
with highest weights $\Lambda$ which satisfy precisely condition $(i)$ of the theorem.


If $v_+$ has weight $\nu$ then it cyclically generates an irreducible $L_0$-module
$\hat{V}_0(\nu)$ of the same highest weight. It follows that $\nu$ must occur in the
decomposition of (\ref{equ7}) (that is, $\nu$ must satisfy the betweenness conditions 
$(i)$ of the theorem) else $\hat{V}_0 (\nu)$ would be orthogonal to $W(\tilde{\Lambda})$ in
contradiction to Lemma \ref{maxwtlemma}. This proves part $(i)$ of the theorem.

As to part $(ii)$, let $v_+$ be as above and let $v^\nu_0 \in W(\tilde{\Lambda})$ be the unique (up to scalar
multiples) maximal weight vector in $W(\tilde{\Lambda})$ with highest weight $\nu$. Suppose $0
\neq w_+ \in V(\tilde{\Lambda})$ is another $L$ highest weight vector of the same
highest weight $\nu$. Then from Lemma \ref{maxwtlemma} we must have
\begin{align*}
\langle w_+,v^\nu_0\rangle \neq 0 ~~,~~ \langle v_+,v^\nu_0\rangle \neq 0.
\end{align*}
Now consider the $L$-maximal vector
\begin{align*}
\tilde{v}_+ = \langle w_+,v^\nu_0\rangle v_+ - \langle v_+,v^\nu_0\rangle w_+.
\end{align*}
By construction we have
\begin{align*}
&\langle\tilde{v}_+, v^\nu_0\rangle = 0\\
\Rightarrow \ & \langle\tilde{v}_+, \hat{V}_0(\nu)\rangle  = \langle\tilde{v}_+,
W(\tilde{\Lambda})\rangle = (0)\\
\Rightarrow \ &\tilde{v}_+ = 0
\end{align*}
by Lemma \ref{maxwtlemma}. This proves that
$w_+ = \kappa v_+,$ with 
\begin{align*}
\kappa = \frac{\langle w_+,v^\nu_0\rangle}{\langle v_+,v^\nu_0\rangle},
\end{align*}
and hence we have proved part $(ii)$ of the theorem.
$\blacksquare$ \medskip

For the irreducible $gl(m|n)$-module $V(\Lambda)$ occuring in the irreducible
$gl(m|n+1)$-module $V(\tilde{\Lambda})$, Theorem \ref{TheoremBR} states that the
conditions (\ref{equ6}) must be satisfied by the highest weights.
Therefore the significance of Theorem \ref{TheoremBR} is that the conditions (\ref{equ6})
are necessary, but not sufficient, so we refer to them as {\em branching conditions}. For an
alternative perspective on the branching rule which we believe will give the reader some
deeper insight, see the discussion in Appendix A.

\section{Invariants and their eigenvalues} \label{evals}

Throughout we let $\psi^p$ denote the odd $gl(m|n)$ vector operator $\psi^p =
(-1)^{(p)}E_{p ,m+n+1}$ $(1 \leq p \leq m+n)$ and we let $\phi_p$
denote the odd $gl(m|n)$ contragredient vector operator $\phi_p = (-1)^{(p)}
E_{m+n+1,p}$. 
There are two natural $gl(m|n)$ invariants we can construct from these
vector operators, namely, for $1\leq r\leq m+n$,
\begin{align}
\gamma_r &= (-1)^{(q)}\phi[r]_q \psi[r]^q = (-1)^{(q)}\phi_q 
\psi[r]^q = E_{m+n+1,q}\psi[r]^q,
\label{equ5.1a}\\
\bar{\gamma}_r &= \psi[r]^q \phi[r]_q = \psi^q 
\phi[r]_q=(-1)^{(p)}E_{p,m+n+1}\phi[r]_p,
\label{equ5.1b}
\end{align}
where, as usual, $\psi[r]^q,$ $\phi[r]_q$ denote the shift components of
$\psi^q,\phi_q$ respectively.
To see that these operators do indeed constitute invariants, consider the following,
noting that we only consider $1\leq p,q\leq m+n$: 
\begin{align}
[E_{pq},\gamma_r] &= [E_{pq},E_{m+n+1,s}\psi[r]^s]\nn\\
&= [E_{pq},E_{m+n+1,s}]\psi[r]^s +
(-1)^{((p)+(q))(1+(s))}
E_{m+n+1,s}[E_{pq},\psi[r]^s]\nn\\
&= -\delta^p_s(-1)^{((p)+(q))(1+(s))}
E_{m+n+1,q}\psi[r]^s + (-1)^{((p)+(q))(1+(s))} \delta^s_q 
E_{m+n+1,s}(-1)^{(p)+(q)}\psi[r]^p\nn\\
&= -(-1)^{((p)+1)(q)} E_{m+n+1,q} \psi[r]^p +
(-1)^{((p)+(q))(q)} E_{m+n+1,q} \psi[r]^p = 0.\nn
\end{align}
A similar calculation can be done in order to determine
\begin{align}
[E_{pq},\bar{\gamma}_r] = 0.\nn
\end{align}
Our interest in these invariants stems from the fact that, by analogy with the normal Lie
algebra situation (see e.g. \cite{Gould1978}), their eigenvalues 
determine the squared {\em reduced matrix elements} of the $gl(m|n+1)$
generators $\psi^p$ and $\phi_p$ respectively.  

As motivation, consider the matrix elements of the $\psi^p,$ in a {\em
unitary} (star) representation of $gl(m|n+1)$ \cite{ZhaGou1990}. 
Using a notation reminiscent of \cite{BB1964}, and keeping in mind
(\ref{eq:SHIFT_COMPONENTS_PSI}) and (\ref{ShiftLeft}), these matrix elements may be
expressed 
$$
\left\langle \left. \left(\begin{array}{c}\Lambda+\varepsilon_r\\ \lambda\end{array}\right)\right|
\psi[r]^p
\left|\left(\begin{array}{c}\Lambda\\ \lambda'\end{array}\right)\right.\right\rangle
=\langle \Lambda+\varepsilon_r||\psi||\Lambda\rangle
\left\langle\left. \left(\begin{array}{c}\Lambda+\varepsilon_r\\ \lambda\end{array}\right)\right|
e_p\otimes \left(\begin{array}{c}\Lambda\\ \lambda'\end{array}\right)\right\rangle,
$$
where $\left|\left.\left(\begin{array}{c}\Lambda\\\lambda'\end{array}\right)\right.\right\rangle$ 
denotes a suitable orthonormal basis for the $gl(m|n)$ module
$V(\Lambda)~\subseteq~V(\tilde{\Lambda})$ concerned, $\{e_p\}$ is a basis for the 
vector representation, and $\langle
\Lambda+\varepsilon_r||\psi||\Lambda\rangle$ is the reduced matrix element. 
Furthermore, let $V(\tilde{\Lambda})$ be a unitary representation such that
$$
\psi^\dagger[r]^p = \phi[r]_p,
$$
from which we obtain
\begin{align}
\left\langle\left. \left(\begin{array}{c}\Lambda\\ \lambda'\end{array}\right) \right|\phi[r]_q\psi[r]^p
\left| \left(\begin{array}{c} \Lambda\\\lambda'' \end{array}\right)  \right.\right\rangle
&=
\left| \langle \Lambda+\varepsilon_r||\psi||\Lambda\rangle \right|^2\nn\\
&\quad \times
\sum_\lambda
\left\langle \left. \left( \begin{array}{c}\Lambda\\ \lambda'\end{array}\right)\otimes
e_q\right|
\left(\begin{array}{c}\Lambda+\varepsilon_r\\ \lambda\end{array}\right) \right\rangle
\left\langle \left. \left( \begin{array}{c}\Lambda+\varepsilon_r\\ \lambda\end{array} \right) \right|
e_p \otimes \left( \begin{array}{c}\Lambda\\ \lambda''\end{array}\right) \right\rangle\nn\\
&= \left| \langle \Lambda+\varepsilon_r||\psi||\Lambda\rangle \right|^2
\left\langle \left. \left( \begin{array}{c}\Lambda\\ \lambda'\end{array}\right) \right|
\bar{P}[r]_q^{\ p}\left| \left( \begin{array}{c} \Lambda\\ \lambda''\end{array} \right)
\right. \right\rangle.
\label{equA1}
\end{align}

At this point we find it instructive to introduce the $gl(m|n+1)$ matrices
\begin{align}
{\cal B}^p_{\ q} &= (-1)^{(p)} E_{pq},\label{equdefy} \\
{\cal \bar{B}}_p^{\ q} &=
-(-1)^{(p)(q)} E_{qp},\ \ 1\leq p,q\leq m+n+1\nn
\end{align}
and the corresponding $gl(m|n+1)$ characteristic roots $\bar{\beta}_r$ and $\beta_r$
($1\leq r\leq m+n+1$), determined by analogy with equations
(\ref{eq:adjCHAR_ROOTSa}), (\ref{eq:adjCHAR_ROOTSb}), (\ref{eq:CHAR_ROOTSa}) and
(\ref{eq:CHAR_ROOTSb}), and which evaluate on an irreducible $gl(m|n+1)$-module
$V(\tilde{\Lambda})$ to
\begin{align}
\bar{\beta}_i &= i-1-\tilde{\Lambda}_i,\nn\\ 
\bar{\beta}_\mu &= \tilde{\Lambda}_\mu+m+1-\mu,\nn\\ 
{\beta}_i &= \tilde{\Lambda}_i+m-n-1-i,\nn\\ 
{\beta}_\mu &= \mu-\tilde{\Lambda}_\mu-n-1,\nn 
\end{align}
with $1\leq i\leq m$ and $1\leq \mu\leq n+1$.
We then also have the associated projection operators
\begin{align}
Q[r] = \prod_{k\neq r}^{m+n+1}\left( \frac{
{\cal B}-\beta_k}{\beta_r-\beta_k} \right),
\ \ 
\bar{Q}[r] = \prod_{k\neq r}^{m+n+1}\left( \frac{{\cal \bar{B}}-\bar{\beta}_k}{\bar{\beta}_r -
\bar{\beta}_k} \right).\nn
\end{align}

It is our aim here to evaluate
the invariants (\ref{equ5.1a}) and (\ref{equ5.1b}) as rational polynomial functions in
the $gl(m|n)$ and $gl(m|n+1)$ characteristic roots.

Note first that the $(m+n+1,m+n+1)$ entries of the projection matrices,
\begin{align}
c_r = Q[r]^{m+n+1}_{\ \ \ \ \ \ \ m+n+1},\ \ \bar{c}_r = \bar{Q}[r]_{m+n+1}^{\ \ \ \ \ \ \ m+n+1},\ \ 1\leq r\leq m+n+1
\label{equ5.2}
\end{align}
clearly determine (even) invariants of $gl(m|n)$ whose eigenvalues, by analogy
with the Lie algebra situation \cite{Gould1981}, are given by certain
$gl(m|n+1)\supset gl(m|n)$ {\em reduced Wigner coefficients}. 

To demonstrate this point by way of example, note that in the context of $gl(m|n+1)$,
using a basis of the unitary $gl(m|n+1)$ module $V(\tilde{\Lambda})$ symmetry adapted to
$gl(m|n)$, we have
\begin{align}
\left\langle\left. \left(\begin{array}{c} \tilde{\Lambda}\\ \Lambda\\ \lambda' \end{array}\right) \right|\right.
\bar{Q}[r]_p^{\ q} \left.\left| \left(\begin{array}{c} \tilde{\Lambda}\\
\Lambda'\\ \lambda'' \end{array}\right)
\right. \right\rangle &=
\sum_{\Lambda'', \lambda} 
\left\langle\left. 
\left(\begin{array}{c} \tilde{\Lambda}\\ \Lambda\\ \lambda'\end{array}\right)
\otimes e_{p} 
\right|\right.
\left.
\left(\begin{array}{c} \tilde{\Lambda}+\varepsilon_r\\ \Lambda''\\ \lambda \end{array}\right)
\right\rangle
\left\langle\left. 
\left(\begin{array}{c} \tilde{\Lambda}+\varepsilon_r\\ \Lambda''\\ \lambda\end{array}\right)
\right|\right.
\left.
e_{q}\otimes \left(\begin{array}{c} \tilde{\Lambda}\\
\Lambda'\\ \lambda'' \end{array}\right)
\right\rangle. \nn
\end{align}
In particular, setting $p=q=m+n+1$ gives
\begin{align}
\left\langle\left. \left(\begin{array}{c} \tilde{\Lambda}\\ \Lambda\\ \lambda' \end{array}\right) \right|\right.
\bar{c}_r \left.\left| \left(\begin{array}{c} \tilde{\Lambda}\\
\Lambda'\\ \lambda'' \end{array}\right)
\right. \right\rangle
=\delta_{\Lambda\Lambda'}\delta_{\lambda'\lambda''}
\left| \left\langle\left. 
\left(\begin{array}{c} \tilde{\Lambda}+\varepsilon_r\\ \Lambda\\ \lambda'\end{array}\right)
\right|\right.
\left.
e_{m+n+1}\otimes \left(\begin{array}{c} \tilde{\Lambda}\\
\Lambda\\ \lambda' \end{array}\right)
\right\rangle \nn \right|^2,
\end{align}
where we have used the fact that $\bar{c}_r$ leaves the representation labels of $gl(m|n)$
unchanged. Thus the invariants $\bar{c}_r$ (and similarly $c_r$) determine squares of
reduced Wigner coefficients. Note that this is the archetypical example where the reduced
Wigner coefficient and Wigner coefficient (c.f. coupling coefficient or Clebsch-Gordan
coefficient) coincide.
%
%
%

%
%

To appreciate the strength of our approach, we demonstrate some results for the
types of invariants we are considering. Recall the recursive definition for powers of the
matrices ${\cal A}$ and ${\cal B}$ respectively, namely 
\begin{align}
\left({\cal A}^k\right)^p_q = \sum_{r=1}^{m+n} {\cal A}^p_r
\left({\cal A}^{k-1}\right)^r_q, ~~\left( \left({\cal A}^0 \right)^p_q \equiv
\delta_{p q} \right), ~~ 1\leq p,q\leq m+n. \label{powerx}
\end{align}
and
\begin{align}
\left({\cal B}^k\right)^p_q = \sum_{r=1}^{m+n+1} {\cal B}^p_r
\left({\cal B}^{k-1}\right)^r_q, ~~\left( \left({\cal B}^0 \right)^p_q \equiv
\delta_{p q} \right), ~~ 1\leq p,q\leq m+n+1.
\label{powery}
\end{align}
We define the two $gl(m|n)$ invariants occuring in $gl(m|n+1)$:
\begin{align}
\tau_k &= ({\cal B}^k)^{m+n+1}_{m+n+1},\nn\\
\sigma_k &= {\cal B}^{m+n+1}_\alpha ({\cal A}^k)^\alpha_\beta {\cal B}^\beta_{m+n+1},\nn
\end{align}
where ${\cal A}$ is the $(m+n) \times (m+n)$ submatrix of ${\cal B}$ that contains the $gl(m|n)$
generators. It is worth observing that the invariants we consider below can then be
expressed in terms of the $\sigma_k$ and $\tau_k$.

Following Green \cite{Green1971} who studied similar objects associated with classical Lie algebras, 
we could adopt the approach of explicitly determining the eigenvalues of these invariants
by attempting to express the invariants themselves in terms of the $gl(m|n+1)$ Casimir
invariants $\hat{I}_k$ and the $gl(m|n)$ Casimir invariant $I_k$ contained in $gl(m|n+1)$,
whose eigenvalues in turn can be expressed in terms of the highest weights labels (see for example
equation (\ref{caseig})). The Casimir invariants are defined by
\begin{align}
\hat{I}_k &= 
(-1)^{(p)}({\cal B}^k)^p_p =  (-1)^{(p)}{\cal B}^p_{q_1}
{\cal B}^{q_1}_{q_2} {\cal B}^{q_2}_{q_3}\ldots {\cal B}^{q_{k-1}}_{p},\ \
1\leq p,q_1,\ldots,q_{k-1}\leq m+n+1, \label{casop1}\\
I_k &= 
(-1)^{(\dot{p})}({\cal A}^k)^{\dot{p}}_{\dot{p}} =
(-1)^{(\dot{p})}{\cal A}^{\dot{p}}_{\dot{q}_1}
{\cal A}^{\dot{q}_1}_{\dot{q}_2} {\cal A}^{\dot{q}_2}_{\dot{q}_3}\ldots
{\cal A}^{\dot{q}_{k-1}}_{\dot{p}},\ \
1\leq\dot{p},\dot{q}_1,\ldots,\dot{q}_{k-1}\leq m+n. \label{casop2}
\end{align}

In Appendix C, we give recursion formulae that enable us to express the $\sigma$ invariants in
terms of the $\tau$ invariants, and more importantly, we are also able to express the
$\tau$
invariants in terms of lower order $\tau$ invariants and the Casimir invariants $I_k$ and
$\hat{I}_k.$ Some examples of the results of such calculations are as follows:
\begin{align}
2 \tau_2 &= \left(I_{2} - \hat{I}_{2} \right) + \left(I_{1} - \hat{I}_{1}  \right)^2 -
I_1 + (m-n)\left(I_{1} - \hat{I}_{1} \right),\nn\\
3 \tau_3 &= \left(I_{3} - \hat{I}_{3} \right) + \tau_1\left(I_{2} - \hat{I}_{2} \right) +
\tau_2\left(I_{1} - \hat{I}_{1} \right) \nn\\
&~~- 2(\tau_2 + \hat{I}_2) - (\tau_1 + \hat{I}_1) + 2(m-n)\tau_2 + (m-n)\tau_1 + \tau_2 +
(\tau_1)^2,\nn\\
4 \tau_4 &= \left(I_{4} - \hat{I}_{4} \right) + \tau_1\left(I_{3} - \hat{I}_{3} \right) +
\tau_2\left(I_{2} - \hat{I}_{2} \right) + \tau_3 \left(I_{1} - \hat{I}_{1} \right) \nn\\
&~~- 3(\hat{I}_3 - (m-n-1)\tau_3) - \hat{I}_2 \tau_1 + \hat{I}_1 \tau_2 + 4\tau_1 \tau_2 -
(\tau_1)^3 \nn\\
&~~+ 3 ( (m-n-1)\tau_2 - \hat{I}_2 + \tau_3) - (\tau_1)^2 + (m-n-1) (\tau_1)^2 - \tau_1 \hat{I}_1  \nn\\
&~~- (m-n-1) \tau_1 + \hat{I}_1 - \tau_2.\nn
\end{align}
See Appendix C for details.

It is clear from these calculations that taking the approach outlined above for determining the
invariants, or more to the point their eigenvalues, leads to 
complicated recursion relations. The results of the present article,
however, completely bypass such complexities, and we find that we are able to present 
elegent eigenvalue formulae using the characteristic identities.

{\bf Remark}: 
\begin{enumerate}
\item[]
Let $\hat{Z}$ be the centre of $U(\hat{L})$ and $Z$ the centre of $U(L)$ for $\hat{L}$
and $L$ Lie superalgebras such that $L\subseteq \hat{L}$. 
In the spirit of Joseph's second commutant theorems \cite{Joseph1977}, we conjecture that
the embedding $gl(m|n)\subset gl(m|n+1)$ is canonical, i.e. that the double commutant of
$L=gl(m|n)$ in $\displaystyle{U\left(\hat{L}=gl(m|n+1)\right)}$ is precisely $U(L)\hat{Z}$. Hence
for this case the centraliser
$$
C(L) = \left\{ u\in U(\hat{L})\ |\ ux-xu=0, \forall x\in L \right\}
$$
of $L$ in $U(\hat{L})$ is given explicitly by $C(L) = \hat{Z}Z$.
\end{enumerate}
%
%
%


From the $gl(m|n+1)$ characteristic identity (i.e. the $gl(m|n+1)$ analogue of
(\ref{eq:POLY_IDENT})), we have ($1\leq p\leq m+n$)
\begin{align}
\sum_{q=1}^{m+n+1}({\cal B}^p_{\ q} - \beta_r\delta^p_{\ q})Q[r]^q_{\ \gamma}=0,\nn
\end{align}
in particular
\begin{align}
\sum_{q=1}^{m+n}{\cal B}^p_{\ q} Q[r]^q_{\ m+n+1} + {\cal B}^p_{\ m+n+1}c_r = \beta_rQ[r]^p_{\ m+n+1},
\nn
\end{align}
which may be rearranged to give 
\begin{align}
\psi^p c_r = \sum_{q=1}^{m+n}(\beta_r - {\cal A})^p_{\ q} Q[r]^q_{\ m+n+1}
\label{equ5.3}
\end{align}
where ${\cal A}$ is the $gl(m|n)$ matrix and we have employed the result
\begin{align}
{\cal B}^p_{\ m+n+1} = (-1)^{(p)} E_{p,m+n+1} = \psi^p,\ \ 1\leq p\leq
m+n.\nn
\end{align}

We note that the $gl(m|n+1)\downarrow gl(m|n)$ branching rules imply that
there may exist degeneracies between the even roots of $gl(m|n+1)$ and those of $gl(m|n).$
Indeed, as we pointed out when the $gl(m|n+1)$ characteristic roots were introduced, the even $gl(m|n+1)$ 
roots are expressible in terms of the $gl(m|n+1)$ representation
labels $\tilde{\Lambda}_i$ as
\begin{align}
\beta_i = \tilde{\Lambda}_i+m-n-1-i,\ \ 1\leq i\leq m.
\nn
\end{align}
In view of the betweenness conditions (\ref{equ6}) we thus have
\begin{align}
\beta_i = \left\{ \begin{array}{rl} \alpha_i,& \tilde{\Lambda}_i = 1+\Lambda_i\\
                                    \alpha_i - 1,& \tilde{\Lambda}_i = \Lambda_i 
\end{array} \right.
\label{equ5.4}\end{align} 
This suggests that we introduce the even index sets
\begin{align}
I_0 = \{ 1\leq i\leq m\ |\ \alpha_i = \beta_i\}, \ \ \bar{I}_0 = \{ 1\leq i\leq m\ |\
\alpha_i = 1+\beta_i\}
\nn
\end{align}
and the full index sets
\begin{align}
I=I_0\cup I_1,\ \ \tilde{I} = I\cup\{ m+n+1\}\nn
\end{align}
where $I_1$ denotes the set of odd indices $\mu=1,2,\ldots, n.$

The importance of introducing the above index sets lies in the fact that if $i\in
\bar{I}_0$ then the shift components $\psi[i]^p$ must vanish on the representation of
$gl(m|n)$ concerned, since the label $\Lambda_i$, $i\in \bar{I}_0$, already takes its
maximum value. In a similar way, the operator $c_i$, $i\in \bar{I}_0,$ must vanish. On the
other hand, for $i\in I_0$, the operators $\phi[i]_p$ and $\bar{c}_i$ must vanish on
the representation concerned. We note also that the operators $Q[i]^p_{\ m+n+1}$ ($i\in
\bar{I}_0$) and $\bar{Q}[i]_p^{\ m+n+1}$ ($i\in I_0$) must vanish on the representation
of $gl(m|n)\subset gl(m|n+1)$ concerned by an analogous argument.

Now inverting equation~(\ref{equ5.3}) 
gives
\begin{align}
Q[r]^p_{\ m+n+1} &= \sum_{q=1}^{m+n}\left( (\beta_r-{\cal A})^{-1} \right)^p_{\
q}\psi^q c_r\nn\\
&= \sum_{q=1}^{m+n}\sum_{r=1}^{m+n}(\beta_r-\alpha_r)^{-1} P[r]^p_{\ q}\psi^q c_r\nn\\
&=\sum_{r=1}^{m+n}(\beta_r-\alpha_r)^{-1} \psi[r]^p c_r\nn\\
&=\sum_{s\in I}(\beta_r - \alpha_s)^{-1}\psi[s]^p c_r,\ \ r\in
\tilde{I}\nn
\end{align}
where we have used the fact that $\psi[s]^p$, $Q[r]^p_{\ m+n+1}$ and $c_r$ all
vanish for $r,s\in\bar{I}_0$, and where $(\beta_r - {\cal A})^{-1}$ denotes the matrix
$$
(\beta_r - {\cal A})^{-1} = \sum_{r=1}^{m+n}(\beta_r-\alpha_r)^{-1} P[r].
$$
It is straightforward to establish the shift relation 
$$
(\beta_r-\alpha_s)^{-1}\psi[s] = \psi[s](\beta_r-\alpha_s-(-1)^{(s)})^{-1},
$$
which then allows us to write 
\begin{align}
Q[r]^p_{\ m+n+1} = \sum_{s\in I} \psi[s]^p(\beta_r-\alpha_s-(-1)^{(s)})^{-1}c_r,\
\ r\in \tilde{I}.
\label{equ5.7}
\end{align}
Summing this equation over $r$ we thus obtain
\begin{align}
\sum_{s\in I}\psi[s]^p \sum_{r\in\tilde{I}} (\beta_r-\alpha_s-(-1)^{(s)})^{-1} c_r =
0
\label{starpage18}
\end{align}
which is a direct consequence of the identity resolution
$$
\sum_{r=1}^{m+n+1} Q[r]^p_{\ m+n+1} = \sum_{r\in\tilde{I}} Q[r]^p_{\ m+n+1} =
\delta^p_{\ m+n+1} = 0, \ \ 1\leq p\leq m+n.
$$
Resolving equation~(\ref{starpage18}) into shift components we obtain, in view of the
linear independence of the $\psi[s]^p$, the following set of equations:
\begin{align}
\sum_{r\in \tilde{I}} \left(\beta_r-\alpha_s-(-1)^{(s)}\right)^{-1}c_r=0,\ \ s\in I.
\label{equ5.8}
\end{align}
Equation (\ref{equ5.8}) yields $|I|$ relations in $|\tilde{I}|=1+|I|$ unknowns $c_r$. To
uniquely determine the $c_r$ we need the extra relation
\begin{align}
\sum_{r\in\tilde{I}}c_r=1
\label{equ5.9}
\end{align}
which follows from the identity resolution
$$
\sum_{r=1}^{m+n+1} Q[r]^{m+n+1}_{\ \ \ \ \ \ \ m+n+1} = \sum_{r\in\tilde{I}}
Q[r]^{m+n+1}_{\ \ \ \ \ \ \ m+n+1} =
\delta^{m+n+1}_{\ \ \ \ \ \ \ m+n+1} = 1.
$$
Using straightforward techniques of linear algebra, equations (\ref{equ5.8}) and
(\ref{equ5.9}) yield the unique solution
\begin{align}
c_s = \prod_{k\in\tilde{I},k\neq s} (\beta_s - \beta_k)^{-1}\prod_{r\in I}(\beta_s -
\alpha_r-(-1)^{(r)}),\ \ s\in \tilde{I}
\label{equ5.10}
\end{align}
which may be expressed in terms of odd and even indices according to
\begin{align}
c_i &= 0,\ \ i\in\bar{I}_0,\nn\\
c_i &= -\prod_{j\in I_0,j\neq i} \left( \frac{\alpha_i-\alpha_j-1}{\alpha_i-\alpha_j} \right)
\prod_{\mu=1}^{n+1}(\beta_i-\beta_\mu)^{-1}\prod_{\mu=1}^n(\beta_i-\alpha_\mu+1),\ \ i\in
I_0,\nn\\
c_\mu &= \prod_{i\in I_0}\left( \frac{\beta_\mu-\beta_i-1}{\beta_\mu-\beta_i} \right)
\prod_{\nu\neq\mu}^{n+1}(\beta_\mu-\beta_\nu)^{-1}\prod_{\nu=1}^n(\beta_\mu-\alpha_\nu+1),
\ \ 1\leq\mu\leq n+1.
\nn
\end{align}
We note that these formulae can be expressed independently of the index set notation as
\begin{align}
c_i &= (\alpha_i-\beta_i-1)
\prod_{k\neq i}^m\left( \frac{\beta_i-\beta_k-1}{\beta_i-\alpha_k} \right)
\prod_{\nu=1}^{n+1}(\beta_i-\beta_\nu)^{-1}\prod_{\nu=1}^n(\beta_i-\alpha_\nu+1),\ \ 1\leq
i\leq m,\nn\\
c_\mu &= \prod_{k=1}^m\left( \frac{\beta_\mu-\beta_k-1}{\beta_\mu-\alpha_k} \right)
\prod_{\nu\neq \mu}^{n+1}(\beta_\mu - \beta_\nu)^{-1} \prod_{\nu=1}^n (\beta_\mu -
\alpha_\nu+1),\ \ 1\leq \mu\leq n+1.\nn
\end{align}

Equation~(\ref{equ5.10}) enables a uniform evaluation of all $gl(m|n)$
invariants of the form 
$$
p({\cal B})^{m+n+1}_{\ \ \ \ \ \ \ m+n+1},
$$ 
for arbitrary polynomials $p(x)$, according to the spectral resolution
$$
p({\cal B})^{m+n+1}_{\ \ \ \ \ \ \ m+n+1} = \sum_{r\in\tilde{I}} p(\beta_r)c_r.
$$

Using the $gl(m|n+1)$ adjoint identity we similarly have
$$
{\cal \bar{A}}_p^{\ q} \bar{Q}[r]_q^{\ m+n+1} + \bar{{\cal B}}_p^{\ m+n+1} \bar{c}_r =
\bar{\beta}_r\bar{Q}[r]_p^{\ m+n+1} 
$$
which may be rearranged to give
$$
-\phi_p\bar{c}_r = (\bar{\beta}_r - {\cal \bar{A}})_p^{\ q}
\bar{Q}[r]_q^{\ m+n+1},
$$
where we have used ${\cal \bar{B}}_p^{\ m+n+1} = -(-1)^{(p)} E_{m+n+1,p} =
-\phi_p$. 
Inverting this equation and resolving $\phi_p$ into its shift
components as before, noting in this case that
$$
\phi[i]_p = \bar{Q}[i]_p^{\ m+n+1} = 0,\ \ i\in I_0,
$$
we obtain
\begin{align}
\bar{Q}[r]_p^{\ m+n+1} = -\sum_{s\in I'}\phi[s]_p(\bar{\beta}_r- \bar{\alpha}_s -
(-1)^{(s)})^{-1} \bar{c}_r,\ r\in \tilde{I}'
\label{equ5.11}
\end{align}
where $\tilde{I}' = I'\cup \{ m+n+1\}$ denotes the index set given by $I' = \bar{I}_0\cup
I_1$ (note that $I'\cap I = I_1$ is the set of odd indices). 
In this case, we obtain by
analogy with equation~(\ref{equ5.10}), 
\begin{align}
\bar{c}_s = \prod_{k\in \tilde{I}',k\neq s} \left(\bar{\beta}_s - \bar{\beta}_k\right)^{-1}\prod_{r\in
I'} \left(\bar{\beta}_s - \bar{\alpha}_r - (-1)^{(r)}\right),\ \ s\in \tilde{I}'
\label{equ5.12}
\end{align}
and, of course, $\bar{c}_s=0$ for $s\in I_0$. 
As in the case of the $c_s$, we can express these formulae independently of the index set notation as
\begin{align}
\bar{c}_i &= (\beta_i-\alpha_i)\prod_{k\neq i}^m\left( \frac{\beta_k-\beta_i-1}{\alpha_k-\beta_i-1} \right)
\prod_{\nu=1}^{n+1}(\beta_\nu - \beta_i-2)^{-1}
\prod_{\nu=1}^n(\alpha_\nu-\beta_i-2),\ \ 1\leq i\leq m,\nn\\
\bar{c}_\mu &= \prod_{k=1}^m\left( \frac{\beta_k-\beta_\mu+1}{\alpha_k-\beta_\mu+1} \right)
\prod_{\nu\neq\mu}^{n+1}(\beta_\nu-\beta_\mu)^{-1}
\prod_{\nu=1}^n(\alpha_\nu -\beta_\mu),\ \ 1\leq\mu\leq n+1.\nn
\end{align}
We summarise the results in the following theorem.

\begin{thm} \label{TheoremC}
The $gl(m|n)$ invariants $c_s$ and $\bar{c}_s$, as given in 
(\ref{equ5.2}), have eigenvalues on an irreducible $gl(m|n)$
module, with highest weight subject to the branching conditions (\ref{equ6}), given
respectively by equations (\ref{equ5.10}) and (\ref{equ5.12}).
\end{thm}

To evaluate the invariants $\gamma_r$ of equation~(\ref{equ5.1a}), we invert
equation~(\ref{equ5.7}) to obtain $\psi[r]^p$ ($r\in I$) as a linear combination of
the $Q[s]^{p}_{\ m+n+1}$ ($s\in\tilde{I}$).
This leads us to look for the unique solution
$\gamma_{rs}$ ($s\in\tilde{I}$, $r\in I$) to the set of equations
\begin{align}
& \sum_{s\in\tilde{I}}\gamma_{rs}\left(\beta_s - \alpha_q - (-1)^{(q)}\right)^{-1} c_s = \delta_{rq},\
\ r,q\in I,
\label{equ5.13a}\\
& \sum_{s\in\tilde{I}}\gamma_{rs}c_s = 0.\label{equ5.13b}
\end{align}
Then for each $r\in I$, equations (\ref{equ5.13a}) and (\ref{equ5.13b}) yield
$|\tilde{I}| = |I|+1$ equations in $|\tilde{I}|$ unknowns $\gamma_{rs}$ ($s\in
\tilde{I}$). These equations are easily solved using standard techniques of linear algebra
and yield the unique solution
$$
\gamma_{rs} = -\gamma_r\left(\beta_s - \alpha_r - (-1)^{(r)}\right) ^{-1},\ \ r\in I,\ \ s\in
\tilde{I},
$$
where
\begin{align}
\gamma_r = (-1)^{|\tilde{I}|} \prod_{q\in I,q\neq r} \left(\alpha_r - \alpha_q + (-1)^{(r)} -
(-1)^{(q)}\right)^{-1}\prod_{p\in\tilde{I}} \left(\beta_p - \alpha_r - (-1)^{(r)}\right).
\label{equ5.14}
\end{align}
As the notation above suggests, formula (\ref{equ5.14}) determines the eigenvalues of the
invariants (\ref{equ5.1a}), as we shall now demonstrate. Multiplying
equation~(\ref{equ5.7}) on the right by $\gamma_{qr}$, summing on $r\in \tilde{I}$ and
making use of equation (\ref{equ5.13a}) we have
\begin{align}
\sum_{r\in\tilde{I}}Q[r]^p_{\ m+n+1}\gamma_{qr} &= \sum_{s\in I}\psi[s]^p 
\sum_{r\in\tilde{I}} \gamma_{qr}\left( \beta_r - \alpha_s - (-1)^{(s)}
\right)^{-1}c_r\nn\\
&= \psi[q]^p.\nn
\end{align}
Thus we obtain
\begin{align}
(-1)^{(p)}\phi[q]_p\psi[q]^p
&= (-1)^{(p)}\phi_p\psi[q]^p\nn\\
&= \sum_{r\in\tilde{I}} (-1)^{(p)}\phi_p Q[r]^p_{\ m+n+1} \gamma_{qr}\nn\\
&= -\sum_{r\in\tilde{I}} \left( \beta_r - {\cal B}^{m+n+1}_{\ \ \ \ \ \ \ m+n+1} \right) c_r\gamma_{qr}\nn\\
&= -\sum_{r\in\tilde{I}} \beta_r c_r \gamma_{qr}\nn
\end{align}
where we have used the result $(-1)^{(p)}\phi_p = E_{m+n+1,p} =
-{\cal B}^{m+n+1}_{\ \ \ \ \ \ \ p}$ and employed equation (\ref{equ5.13b}) in the last step. Thus we may
write
\begin{align}
(-1)^{(p)}\phi[q]_p\psi[q]^p 
&= -\sum_{r\in \tilde{I}} \beta_rc_r\gamma_{qr}\nn\\
&= \gamma_q\sum_{r\in\tilde{I}} \beta_r c_r \left( \beta_r-\alpha_q - (-1)^{(q)}
\right)^{-1} \nn\\
&= \gamma_q \left( \sum_{r\in\tilde{I}} c_r + \left( \alpha_q+(-1)^{(q)} \right)
\sum_{r\in\tilde{I}} c_r \left( \beta_r - \alpha_q - (-1)^{(q)} \right)^{-1} \right)\nn\\
&= \gamma_q,\nn
\end{align}
where in the last step we employed equations (\ref{equ5.8}) and (\ref{equ5.9}). This shows
the required result, that the eigenvalues of the $gl(m|n)$ invariants
$$
\gamma_q = (-1)^{(p)}\phi[q]_p\psi[q]^p
$$
of equation (\ref{equ5.1a}) are given by the formula (\ref{equ5.14}). 

In a similar way, by employing equation (\ref{equ5.11}), we arrive at the following
formula for the invariants $\bar{\gamma}_q = \psi[q]^p\phi[q]_p$ of equation
(\ref{equ5.1b}):
\begin{align}
\bar{\gamma}_q = (-1)^{|I'|} \prod_{r\in I',r\neq q}\left( \bar{\alpha}_q - \bar{\alpha}_r +
(-1)^{(q)} - (-1)^{(r)} \right)^{-1} \prod_{s\in\tilde{I}'} \left( \bar{\beta}_s -
\bar{\alpha}_q - (-1)^{(q)} \right),
\label{equ5.15}
\end{align}
where the index sets $I',\tilde{I}'$ are as in equation~(\ref{equ5.12}). 
Note that for
$i\in I_0$, $\bar{\gamma}_i=0$, while for $i\in \bar{I}_0,$ $\gamma_i=0.$ The remaining
non-zero cases are given by equations (\ref{equ5.14}) and (\ref{equ5.15}). 
Using the easily established relations
$$
\alpha_s+\bar{\alpha}_s = m-n-(-1)^{(s)},\ \ \beta_s + \bar{\beta}_s = m-n-1-(-1)^{(s)},
$$
we note that equation~(\ref{equ5.15}) may be alternatively expressed
\begin{align}
\bar{\gamma}_q = (-1)^{|I'|}\prod_{r\in I',r\neq q}(\alpha_q -
\alpha_r)^{-1}\prod_{s\in\tilde{I}'} \left( \beta_s - \alpha_q + (-1)^{(s)}+1 \right),\ \
q\in I'.
\label{equ5.15d}
\end{align}
In graded index notation, we thus obtain the following formulae for the invariants~$\gamma_r,\bar{\gamma}_r$:
\begin{align}
\gamma_i &= 0,\ \ i\in \bar{I}_0,\nn\\
\gamma_i &= \prod_{j\in I_0,j\neq i}\left( \frac{\alpha_i-\alpha_j+1}{\alpha_i-\alpha_j} \right)
\prod_{\mu=1}^n(\alpha_i-\alpha_\mu+2)^{-1} \prod_{\mu = 1}^{n+1}(\alpha_i -
\beta_\mu+1),\ \ i\in I_0,\nn\\
\gamma_\mu & = \prod_{j\in I_0}\left( \frac{\alpha_\mu-\alpha_j-1}{\alpha_\mu-\alpha_j} \right)
\prod_{\nu\neq\mu}^n(\alpha_\mu-\alpha_\nu)^{-1}\prod_{\nu=1}^{n+1}(\alpha_\mu-\beta_\nu-1),\
\ 1\leq \mu\leq n,\nn\\ 
\bar{\gamma}_i &= 0,\ \ i\in I_0,\nn\\
\bar{\gamma}_i &= \prod_{j\in \bar{I}_0,j\neq i}\left( \frac{\alpha_i-\alpha_j-1}{\alpha_i-\alpha_j} \right)
\prod_{\mu=1}^n(\alpha_i-\alpha_\mu)^{-1} \prod_{\mu = 1}^{n+1}(\alpha_i -
\beta_\mu),\ \ i\in \bar{I}_0,\nn\\
\bar{\gamma}_\mu &= \prod_{j\in \bar{I}_0}\left( \frac{\alpha_\mu-\alpha_j-1}{\alpha_\mu-\alpha_j} \right)
\prod_{\nu\neq\mu}^n(\alpha_\mu-\alpha_\nu)^{-1}\prod_{\nu=1}^{n+1}(\alpha_\mu-\beta_\nu),\
\ 1\leq \mu\leq n.\nn
\end{align}
Alternatively, we may express these formulae independently of the index sets:
\begin{align}
\gamma_i &= (\beta_i-\alpha_i+1)
\prod_{k\neq i}^m\left( \frac{\alpha_k-\alpha_i-1}{\beta_k-\alpha_i} \right)
\prod_{\nu=1}^{n+1}(\beta_\nu-\alpha_i-1)\prod_{\nu=1}^n(\alpha_\nu-\alpha_i-2)^{-1},\ \
1\leq i\leq m,\nn\\
\gamma_\mu &= \prod_{k=1}^m\left(
\frac{\alpha_k-\alpha_\mu+1}{\beta_k-\alpha_\mu+2} \right)
\prod_{\nu=1}^{n+1}(\beta_\nu-\alpha_\mu+1)
\prod_{\nu\neq\mu}^n(\alpha_\nu-\alpha_\mu)^{-1},\ \ 1\leq\mu\leq n,\nn\\
\bar{\gamma}_i &= (\alpha_i-\beta_i)
\prod_{k\neq i}^m\left( \frac{\alpha_k-\alpha_i+1}{\beta_k-\alpha_i+1} \right)
\prod_{\nu=1}^{n+1} (\beta_\nu-\alpha_i)
\prod_{\nu=1}^n(\alpha_\nu-\alpha_i)^{-1},\ \ 1\leq i\leq m,   \nn\\
\bar{\gamma}_\mu &= -\prod_{k=1}^m\left(
\frac{\alpha_k-\alpha_\mu+1}{\beta_k-\alpha_\mu+1} \right)
\prod_{\nu=1}^{n+1}(\beta_\nu-\alpha_\mu)
\prod_{\nu\neq\mu}^n(\alpha_\nu-\alpha_\mu)^{-1},\ \ 1\leq \mu\leq n.  \nn
\end{align}
We summarise the previous discussion in the following theorem, which is analogous to
Theorem \ref{TheoremC}, but for $\gamma_r$ and $\bar{\gamma}_r$.

\begin{thm} \label{TheoremGAMMA}
The $gl(m|n)$ invariants $\gamma_r$ and $\bar{\gamma}_r$, as given by
equations (\ref{equ5.1a}) and (\ref{equ5.1b}) respectively, have
eigenvalues on an irreducible $gl(m|n)$
module, with highest weight subject to the branching conditions (\ref{equ6}), that are
either zero or given respectively by equations (\ref{equ5.14}) and (\ref{equ5.15}).
\end{thm}

We now note, as in the corresponding Lie algebra case \cite{Gould1978}, that
\begin{align}
(-1)^{(q)}\psi[r]^p \left( \gamma_r \right)^{-1}\phi[r]_q &=
P[r]^p_{\ q},\ \ r\in I',\label{equ5.18a}\\
\phi[r]_p \left( \bar{\gamma}_r \right)^{-1}\psi[r]^q &=
\bar{P}[r]_p^{\ q},\ \ r\in I. \label{equ5.18b}
\end{align} 
The above suggests that we define new invariants $\delta_r,\bar{\delta}_r$ such that
\begin{align}
\delta_r\psi[r] = \psi[r] \gamma_r,\ \ \bar{\delta}_r \phi[r] = \phi[r]\bar{\gamma}_r,
\label{deltas}
\end{align}
in which case equations (\ref{equ5.18a}) and (\ref{equ5.18b}) may be rewritten
\begin{align}
(-1)^{(q)}\psi[r]^p\phi[r]_q &= \delta_r P[r]^p_{\ q},\ \ r\in
I',\label{equ5.19a}\\
\phi[r]_p \psi[r]^q &= \bar{\delta}_r \bar{P}[r]_p^{\ q},\ \ r\in I.
\label{equ5.19b}
\end{align}
To justify the relevance of these equations to reduced matrix elements, 
compare equation~(\ref{equ5.19b}) to equation~(\ref{equA1}) in the case of unitary
representations. In such a case, this shows that the $gl(m|n)$ invariants $\bar{\delta}_r$
determine the squared reduced matrix elements of $\psi[r]^p$. We can apply an
analogous argument to show the invariants $\delta_r$ determine the squared reduced matrix
elements of $\phi[r]_p$. Even in the non-unitary case, we expect these invariants
will still determine squared reduced matrix elements. We may therefore be inclined to refer to
$\delta_r,\bar{\delta}_r$ as {\em generalised squared reduced matrix elements}. 

Using our formulae for $\gamma_r,\bar{\gamma}_r$, we easily obtain the following corollary
to Theorem \ref{TheoremGAMMA}.
\begin{cor} \label{CorollaryDELTA}
The $gl(m|n)$ invariants $\delta_r$ and $\bar{\delta}_r$,
satisfying (\ref{deltas}), have eigenvalues on an irreducible $gl(m|n)$
module, with highest weight subject to the branching conditions (\ref{equ6}), given by
\begin{align}
\delta_r &= (-1)^{|I|} \prod_{s\in I,s\neq r} \left(\alpha_r - \alpha_s -
(-1)^{(s)}\right)^{-1}\prod_{q\in\tilde{I}} \left(\beta_q - \alpha_r \right),\ \ r\in I',
\label{equ5.20}\\
\bar{\delta}_r &= (-1)^{|I'|} \prod_{q\in I',q\neq r}\left( \alpha_r-\alpha_q+(-1)^{(r)}
\right)^{-1} \prod_{s\in \tilde{I}'}\left( \beta_s - \alpha_r + (-1)^{(s)} - (-1)^{(r)}+1
\right), r\in I. \label{equ5.21}
\end{align}
\end{cor}

We remark that the formulae (\ref{equ5.20}) and (\ref{equ5.21}) of Corollary
\ref{CorollaryDELTA} may be expressed independently of the index set notation as
\begin{align}
\delta_i &= (\beta_i-\alpha_i)
\prod_{k\neq i}^m \left( \frac{\alpha_k-\alpha_i}{\beta_k-\alpha_i+1} \right)
\prod_{\nu=1}^n(\alpha_\nu-\alpha_i-1)^{-1}
\prod_{\nu=1}^{n+1}(\beta_\nu-\alpha_i),\ \ 1\leq i\leq m,\nn\\
\delta_\mu &= -\prod_{k=1}^m\left(
\frac{\alpha_k-\alpha_\mu}{\beta_k-\alpha_\mu+1} \right)
\prod_{\nu\neq\mu}^n(\alpha_\nu-\alpha_\mu-1)^{-1}
\prod_{\nu=1}^{n+1}(\beta_\nu-\alpha_\mu),\ \ 1\leq \mu\leq n,\nn\\
\bar{\delta}_i &= (\beta_i-\alpha_i+1) \prod_{k\neq
i}^m\left(\frac{\alpha_k-\alpha_i}{\beta_k-\alpha_i}  \right)
\prod_{\nu=1}^{n+1}(\beta_\nu-\alpha_\mu-1)
\prod_{\nu=1}^n(\alpha_\nu-\alpha_i-1)^{-1},\ \ 1\leq i\leq m,\nn\\
\bar{\delta}_\mu &= -\prod_{k=1}^m\left(
\frac{\alpha_k-\alpha_\mu+2}{\beta_k-\alpha_\mu+2} \right)
\prod_{\nu=1}^{n+1}(\beta_\nu-\alpha_\mu+1)
\prod_{\nu\neq\mu}^n(\alpha_\nu-\alpha_\mu+1)^{-1},\ \ 1\leq\mu\leq n.\nn
\end{align}

As a check on these equations, we note that taking the supertrace of equations (\ref{equ5.19a})
and (\ref{equ5.19b}) shows that the invariants discussed are related by
\begin{align}
\bar{\gamma}_r &= \delta_r \mbox{str}(P[r]),\label{equ5.22a}\\
\gamma_r &= \bar{\delta}_r \mbox{str}(\bar{P}[r]),\label{equ5.22b}
\end{align}
which can be regarded as defining $\delta_r,\bar{\delta}_r$. We note immediately that 
$$
\delta_i=0,\ \ i\in I_0,\ \ \bar{\delta}_i=0,\ \ i\in \bar{I}_0,
$$
so that equations (\ref{equ5.20}) and (\ref{equ5.21}) give all the non-zero eigenvalues of
the invariants $\delta_r,\bar{\delta}_r$.

Our interest here is the fact that equations (\ref{equ5.22a}) and (\ref{equ5.22b}) enable
us to evaluate the supertraces of the $gl(m|n)$ projection operators and hence the
eigenvalues of all $gl(m|n)$ Casimir invariants, previously obtained by Jarvis and Green
\cite{JarGre1979} using other methods. We have
$$
\mbox{str}(P[r]) = \bar{\gamma}_r\delta_r^{-1}.
$$
Note that strictly speaking, this formula is only valid on the Zariski dense subset
as mentioned earlier. For more on the general case, see \cite{GouldLinks1996,GouldLinks1997}.
In graded index notation, we furthermore have
\begin{align}
\bar{\gamma}_r &= (-1)^{|I'|} \prod_{i\in \bar{I}_0,i\neq r}(\alpha_r - \alpha_i)^{-1}
\prod_{i\in \bar{I}_0}(\alpha_i+1-\alpha_r)\prod_{\nu\neq r}^n(\alpha_r-\alpha_\nu)^{-1}
\prod_{\nu=1}^{n+1}(\beta_\nu - \alpha_r)\nn\\
\left( \delta_r\right)^{-1} 
&= (-1)^{|I|}\prod_{i\in I_0}(\alpha_r-\alpha_i - 1)\prod_{i\in
I_0}(\alpha_i-\alpha_r)^{-1}\prod_{\nu\neq
r}^n(\alpha_r-\alpha_\nu+1)\prod_{\nu=1}^{n+1}(\beta_\nu - \alpha_r)^{-1}\nn
\end{align}
\begin{align}
\Rightarrow\ \ \mbox{str}(P[r]) = \prod_{q\neq r}^{m+n}\left( \frac{\alpha_r - \alpha_q -
(-1)^{(q)}}{\alpha_r - \alpha_q} \right)
\label{equ5.23}
\end{align}
where in the above, we exploited the fact that $\beta_i = \alpha_i$ (respectively
$\alpha_i-1$) for $i\in I_0$ (respectively $\bar{I}_0$). Equation (\ref{equ5.23}) agrees with
the supertrace formula, previously derived by Jarvis and Green \cite{JarGre1979}, which
provides a check on our formalism. Care is needed, however, in dealing with {\em atypical}
representations.

It follows, in view of the spectral resolution, that the eigenvalues of the
superinvariants
$$
I_k = \mbox{str}({\cal A}^k) = \sum_{p=1}^{m+n} (-1)^{(p)} \left( {\cal A}^k
\right)^p_{\ p}
$$
are given by
$$
I_k = \sum_{r=1}^{m+n} (\alpha_r)^k \mbox{str}(P[r]).
$$
In a similar way we have the adjoint invariants
$$
\bar{I}_k=\mbox{str}(\bar{{\cal A}}^k) =
\sum_{r=1}^{m+n}(\bar{\alpha}_r)^k\mbox{str}(\bar{P}[r])
$$
which may be evaluated with the help of the easily established formula
$$
\mbox{str}(\bar{P}[r]) = \gamma_r\left( \bar{\delta}_r \right)^{-1} = 
\prod_{q\neq r}^{m+n} \left(
\frac{\alpha_r-\alpha_q+(-1)^{(r)}}{\alpha_r-\alpha_q+(-1)^{(r)}-(-1)^{(q)}} \right).
$$

\section{Concluding remarks}

In this paper we have demonstrated how characteristic identities and projection techniques
can be applied to determine certain $gl(m|n)$ invariants in (an algebraic extension of) 
the enveloping algebra of $gl(m|n+1)$. We have provided
explicit derivations of the eigenvalues of such invariants on any irreducible
representation, and our key formulae were presented in Theorems \ref{TheoremC} and
\ref{TheoremGAMMA} and Corollary \ref{CorollaryDELTA}. 

The next step will be the application of these results to {\em unitary} irreducible
representations, and the explicit derivation of the matrix elements of generators. 
In fact, in Section \ref{evals} of the current article we mention 
the class of unitary representations as motivation for the current work, although our
results are not restricted to this class. 
Referring to the classification of such irreducible representations given in
\cite{ZhaGou1990}, caution is needed since the tensor product of a type 1 unitary with a
type 2 unitary may contain indecomposables. 
We note that the vector representation is type 1, and its dual is type 2, so based on the
presentation in this article, we would need to present separate treatments of the cases
where $V(\Lambda)$ is type 1 or type 2, although the two are related via duality. 
Indeed, the current article demonstrates how far one can go without making further
restrictions on the class of irreducible representation.


\section*{Appendix A}

To assist in understanding the properties of odd vector and contragredient vector operators,
we now construct the appropriate odd vector representation. A realisation of odd
vector operators can be constructed by introducing a set of \textit{fermion} operators
$$
a_i^\dagger, a_i, \ \ 1 \leq i \leq m
$$ 
and bosons 
$$
b_\mu^\dagger, b_\mu,\ \ 1 \leq \mu \leq n
$$
according to which our $gl(m|n)$ generators are realised by
\begin{align}
E_{ij} = a_i^\dagger a_j,~E_{i\mu} = a_i^\dagger b_\mu,~ E_{\mu i} = b_\mu^\dagger a_i,~
E_{\mu \nu} = b_\mu^\dagger b_\nu\nn
\end{align}
(note the reversal of grading implicit with the description). Then the operators
\begin{align}
\chi^i = a_i^\dagger,~ \chi^\mu = b_\mu^\dagger\nn
\end{align}
transform as an odd vector operator.
We have the rank two tensor
\begin{align}
T^{ij} = a_i^\dagger a_j^\dagger = -T^{ji}, T^{i\mu} = T^{\mu i} = a_i^\dagger
b_\mu^\dagger, T^{\mu \nu} = T^{\nu \mu} = b_\mu^\dagger b_\nu^\dagger\nn
\end{align}
whose components transform as the representation $(1,1,\dot{0})$ of $gl(m|n)$. The
components of the tensor $T$ thus satisfy
\begin{align}
T^{p q} = (-1)^{((p) + 1)((q) + 1)} T^{q p}.
\label{eq:TWO_RANK_TENSOR}
\end{align}
The weights in the representation $(1,1,\dot{0})$ are of the form (in the notation of Kac
\cite{Kac1978})
\begin{align}
\varepsilon_i + \varepsilon_j \ (i \neq j), \ \ \varepsilon_i + \delta_\mu,\ \  \delta_\mu + \delta_\nu.\nn
\end{align}
We wish to apply these results to investigate odd vector operators whose components
commute (in the graded bracket sense).

Hence we now assume that $\psi^p$ is an odd vector operator whose components are
graded-commuting:
\begin{align}
[\psi^p,\psi^q] = 0
\nn
\end{align}
where the graded bracket is given by
\begin{align}
[\psi^p,\psi^q] = \psi^p \psi^q - (-1)^{((p) + 1)((q) + 1)}
\psi^q \psi^p.\nn
\end{align}
In other words, we are assuming that the components of $\psi$ satisfy the symmetry rule
\begin{align}
\psi^p \psi^q = (-1)^{((p) + 1)((q) + 1)} \psi^q \psi^p.
\label{eq:SYMMETRY_RULE}
\end{align}
Equating even shift components of equation~(\ref{eq:SYMMETRY_RULE}) we obtain
\begin{align}
\psi[i]^p \psi[i]^q = (-1)^{((p) + 1)((q) + 1)} \psi[i]^q
\psi[i]^p, ~~ 1 \leq i \leq m.\nn
\end{align}
It follows, in view of equation~(\ref{eq:TWO_RANK_TENSOR}), that
\begin{align}
T^{p q} = \psi[i]^p \psi[i]^q\nn
\end{align}
constitutes a tensor operator of $gl(m|n)$ transforming as the representation $(1, 1,
\dot{0})$ and which increases the representation labels by the weight $2\varepsilon_i$. But
this latter weight is not a weight of $(1, 1, \dot{0})$ so we must have
\begin{align}
\psi[i]^p \psi[i]^q = 0, ~~~ 1 \leq i \leq m.\nn
\end{align}
More generally, by focusing on higher order tensors, we may establish the result
\begin{align}
\ldots \psi[i]^p \psi[t]^r \ldots \psi[u]^s \psi[i]^q \ldots = 0,\ \  1 \leq i \leq m.
\nn
\end{align}
That is, any product of shift components of $\psi$ must vanish if there are two even shift
indices which are equal. This result is a direct consequence of the oddness of $\psi$ and
the fact that its components are graded-commuting.
Similarly, if $\phi_p$ is an odd contragredient vector operator such that
\begin{align}
\phi_p \phi_q = (-1)^{((p) + 1)((q) + 1)} \phi_q \phi_p\nn 
\end{align}
then
\begin{align}
\phi[i]_p \phi[i]_q = 0,~~~ 1 \leq i \leq m\nn
\end{align}
and more generally
\begin{align}
\ldots \phi[i]_p \phi[t]_r \ldots \phi[u]_s \phi[i]_q \ldots = 0,\ \ 1
\leq i \leq m.\nn
\end{align}
It turns out that these results may be applied to provide useful information on the $gl(m|n+1)\downarrow
gl(m|n)$ branching conditions derived in Section \ref{br}.

Let $V(\tilde{\Lambda})$ be an irreducible finite dimensional $gl(m|n+1)$ module with
highest weight
\begin{align}
\tilde{\Lambda} = \sum^m_{i=1} \tilde{\Lambda}_i \varepsilon_i + \sum^{n+1}_{\mu=1}
\tilde{\Lambda}_\mu \delta_\mu = \tilde{\Lambda}_0 + \tilde{\Lambda}_1.
\nn
\end{align}
Let
\begin{align}
V_0(\tilde{\Lambda}) = V(\tilde{\Lambda}_0) \otimes V(\tilde{\Lambda}_1)\nn
\end{align}
be the irreducible $\hat{L}_0=gl(m) \oplus gl(n+1)$ module which has a grading index $0$ under the
usual $\mathbb{Z}$-gradation induced on $V(\tilde{\Lambda})$ by the odd roots, i.e.
\begin{align}
E_{i\mu} V_0 (\tilde{\Lambda}) = 0,~~~ 1 \leq i \leq m, ~ 1 \leq \mu \leq n + 1.\nn
\end{align}
Then this representation decomposes into irreducible representations of $gl(m) \oplus
gl(n)$ according to
\begin{align}
\label{eq:EVEN_DECOMP}
V_0(\tilde{\Lambda}) =  \bigoplus_\Lambda V_0 (\Lambda),~~~ V_0(\Lambda) =
V(\tilde{\Lambda}_0) \otimes V(\Lambda_1)\ \ \ 
\end{align}
where $V_0(\Lambda)$ is an irreducible representation of $gl(m) \oplus gl(n)$ and the sum
is over all $gl(m|n)$ highest weights
\begin{align}
\label{eq:HIGHEST_WEIGHT}
\Lambda = \sum^m_{i=1} \tilde{\Lambda}_i \varepsilon_i + \sum^n_{\mu=1} \Lambda_\mu
\delta_\mu= \tilde{\Lambda}_0 + \Lambda_1
\end{align}
where the odd components $\Lambda_\mu$ are subject to the usual Gelfand-Tsetlin
betweenness conditions \cite{GT1950}:
\begin{align}
 \label{eq:BETWEENESS_COND}
\tilde{\Lambda}_\mu \geq \Lambda_\mu \geq \tilde{\Lambda}_{\mu + 1}, ~~~ 
1 \leq \mu \leq n
\end{align}
Applying the odd $gl(m|n)$ lowering operators $E_{\mu i}$ $(1 \leq i \leq m,~ 1 \leq \mu
\leq n)$ to the left of equation~(\ref{eq:EVEN_DECOMP}) it follows that
\begin{align}
\label{eq:INCLUDES_DECOMP}
V(\tilde{\Lambda}) \supseteq W
\end{align}
where
\begin{align}
W = \underset{\Lambda}{ \bigoplus} V(\Lambda)
\nn\end{align}
where $V(\Lambda)$ is the irreducible module of $gl(m|n)$ with highest
weight~(\ref{eq:HIGHEST_WEIGHT}) and the sum is over all $\Lambda$ subject to the betweenness
conditions~(\ref{eq:BETWEENESS_COND}).

We now note that the $gl(m|n)$ generator
\begin{align}
\phi_p = (-1)^{(p)} E_{m+n+1, p}, ~~~ 1 \leq p \leq m+n\nn
\end{align}
transforms as an odd contragredient vector operator of $gl(m|n)$:
\begin{align}
[E_{p q}, \phi_r] = -(-1)^{(p)((p) + (q))} \delta_{p r} \phi_q,
\nn\end{align}
whose components commute (in the usual graded bracket sense).
Applying the $gl(m|n+1)$ generators $\phi_p$ $(1 \leq p \leq m+n)$ to the left
of equation~(\ref{eq:INCLUDES_DECOMP}) we obtain
\begin{align}
V(\tilde{\Lambda}) = W + \sum^m_{i=1} \phi_i W + \sum_{i \neq j} \phi_i \phi_j W + ... +
\phi_1 \phi_2 ...\phi_m W,\nn
\end{align}
where we have used the fact that the space $V_0(\tilde{\Lambda})$ (c.f.
equation~(\ref{eq:EVEN_DECOMP})) is stable under the action of the even generators
$\phi_\mu$ $(1 \leq \mu \leq n)$.

We note that only the even shift components of $\phi$ make a new contribution since for
$V(\Lambda) \subseteq W$ we have
\begin{align}
\phi[\mu]_i V(\Lambda) \subseteq V(\Lambda - \delta_\mu) \subseteq W.\nn
\end{align}
Thus we have
\begin{align}
V(\tilde{\Lambda}) = W \oplus \sum^m_{i,j} \phi[i]_j W \oplus ... \oplus
\sum_{\pi\in S_m} \phi[1]_{\pi(1)} ... \phi[m]_{\pi(m)} W\nn
\end{align}
where $S_m$ denotes the symmetric group on $m$ elements, and we have used the previously
established result that no two even shift indices can occur in a product of shift
components of an odd contragredient vector operator. 

This produces an alternative perspective to
understanding the $gl(m|n+1) \downarrow gl(m|n)$ branching conditions presented in Section
\ref{br}. While the result of Theorem \ref{TheoremBR} is quite rigorous, the discussion
in this Appendix
merely provides insight into the general branching rule. Indeed, the outcome of this
discussion is a rough sketch of the branching rule, where we have overlooked complications 
such as the possibility that some of 
the $gl(m|n)$ modules occurring may not be irreducible but only indecomposable (or possibly 
even zero).


\section*{Appendix B} 

Recall the characteristic roots
\begin{equation*}
\bar{\alpha}_r = -\frac12 [\chi_{\Lambda + \varepsilon_r} (I_2) + n - m -
\chi_{\Lambda}(I_2)]
\end{equation*}
of the adjoint identity. We make the basic assumption that these roots are distinct or
equivalently the numbers
\begin{equation}
\label{eq:VTVstar}
\chi_{\Lambda+\varepsilon_r} (I_2) \equiv \chi_{\Lambda + \varepsilon_i}(I_2),
\chi_{\Lambda + \delta_\mu}(I_2)
\end{equation}
are all distinct.

Throughout $V = V_0 \oplus V_1$ (usual $\mathbb{Z}$-grading) is the vector module (i.e
$V=V(\varepsilon_1)$). It is our aim to prove the following theorem under the assumption
(\ref{eq:VTVstar}).
\begin{thm} 
$V \otimes V(\Lambda)$ is completely reducible and the allowed highest weights are of the
form $\Lambda+\varepsilon_r$ $(1 \leq r \leq m+n)$ each occurring at most once.
\end{thm}

Throughout $V_0(\Lambda)$ denotes the maximal $\mathbb{Z}$-graded component of $V(\Lambda)$  which constitutes an irreducible $L_0$-module.

Here $L$ denotes the Lie superalgebra $gl(m|n)$ which admits the usual $\mathbb{Z}$-gradation
\begin{equation*}
L=L_- \oplus L_0 \oplus L_+
\end{equation*}
with
\begin{equation*}
L_0 = gl(m) \oplus gl(n).
\end{equation*}

We begin with the following easily established result:
\begin{lemma} \label{Lemma7}
$V \otimes V(\Lambda)$ is cyclically generated as an $L$-module by the $L_0$-submodule $V \otimes V_0(\Lambda)$ and
\begin{equation*}
V \otimes V(\Lambda) = U(L_-) V \otimes V_0(\Lambda).
\end{equation*}
\end{lemma}

Note that we have the following decomposition into irreducible $L_0$-modules:
\begin{equation}
\label{eq:VTV1a}
V_0 \otimes V_0(\Lambda) = \bigoplus_i V_0 (\Lambda + \varepsilon_i)
\end{equation}
\begin{equation}
\label{eq:VTV1b}
V_1 \otimes V_0(\Lambda) = \bigoplus_\mu V_0 (\Lambda + \delta_\mu)
\end{equation}
where each module on the RHS is understood to vanish identically if one of the
corresponding ($\Lambda+\varepsilon_i$) or ($\Lambda + \delta_\mu$) is non-dominant. The
above gives rise to the following decomposition into irreducible $L_0$-submodules for the
cyclic module of Lemma \ref{Lemma7}:
\begin{equation}
\label{eq:VTV2}
V \otimes V_0(\Lambda) = \bigoplus_r V_0(\Lambda + \varepsilon_r).
\end{equation}
Now let $\left\langle~,~\right\rangle$ be the (unique) non-degenerate sesquilinear form on $V(\Lambda)$ satisfying
\begin{equation*}
\left\langle av,w\right\rangle = \left\langle v,a^\dagger w \right\rangle, ~\forall v,w \in V(\Lambda)
\end{equation*}
with $\dagger$ the usual conjugation operation on $L$. We recall \cite{Kac1978} that the corresponding
form $\left\langle~,~\right\rangle$ on the vector module $V$ gives rise to an inner
product. We let $\langle~,\ \rangle$ denote the form induced on $V\otimes V(\Lambda)$. 

The following Lemma will prove useful:

\begin{lemma}
Let $0 \neq v_+ \in V \otimes V(\Lambda)$ be a maximal weight vector. Then $\left\langle
v_+,V \otimes V_0(\Lambda)\right\rangle \neq (0)$.
\end{lemma}
\proof{
Otherwise we would have
\begin{align*}
0 &= \left\langle U(L_+) v_+, V \otimes V_0(\Lambda) \right\rangle\\
&= \left\langle v_+, U(L_-) V \otimes V_0(\Lambda) \right\rangle\\
&= \left\langle v_+, V \otimes V(\Lambda) \right\rangle \mbox{ (from Lemma \ref{Lemma7})}\\
\Rightarrow v_+ &= 0
\end{align*}
since the naturally induced form on $V \otimes V(\Lambda)$ is also non-degenerate.
}

We now aim to prove the following weaker version of the Theorem above, using the Lemma:

\begin{prop} \label{Prop}
Let $v_+ \in V \otimes V(\Lambda)$ be a maximal weight vector of weight $\nu$. Then
\begin{enumerate}
\item[(i)] $\nu \in \{ \Lambda + \varepsilon_r | 1 \leq r \leq m+n\}$
\item[(ii)] $v_+$ is the unique (up to scalar multiples) maximal weight vector in $V \otimes V(\Lambda)$ of weight $\nu$. 
\end{enumerate}
\end{prop}

{\bf Proof:}
From the decomposition (\ref{eq:VTV2}) this is enough to prove part $(i)$ of the
proposition. As to the second part we proceed as in the proof of Theorem \ref{TheoremBR}
and suppose $0 \neq
w_+ \in  V \otimes V(\Lambda)$ is also a maximal weight vector of the same weight $\nu$.
We let $v^\nu_0$ be the $L_0$ maximal vector of the irreducible $L_0$-module $V_0(\nu)
\subseteq V \otimes V_0(\Lambda) $.
Then we observe that
\begin{equation*}
\tilde{v}_+ \equiv \left\langle v^\nu_0,w_+ \right\rangle v_+  - \left\langle v^\nu_0,v_+ \right\rangle w_+
\end{equation*}
is also a maximal vector of weight $\nu$ satisfying
\begin{align*}
\left\langle \tilde{v}_+, V \otimes V_0(\Lambda) \right\rangle = (0)
\ \Rightarrow\ \tilde{v}_+ = 0
\end{align*}
from the Lemma, so 
\begin{equation*}
w_+ = \kappa v_+~,~\kappa = \frac{\left\langle v^\nu_0,w_+ \right\rangle}{\left\langle v^\nu_0,v_+ \right\rangle}
\end{equation*}
which proves the proposition.
$\blacksquare$ \medskip

Now observe from equation (\ref{eq:VTV1a}) that each irreducible $L_0$-module 
$$
V_0(\Lambda+\varepsilon_i)\subseteq V_0\otimes V_0(\Lambda)
$$ 
contains a maximal weight vector of
weight $\Lambda + \varepsilon_i$, which is also a maximal weight vector of $L$ cyclically
generating an indecomposable $L$-module
\begin{equation}
\label{eq:VTV3}
V(\Lambda + \varepsilon_i) = U(L_-) V_0 (\Lambda + \varepsilon_i).
\end{equation}
If $V(\Lambda + \varepsilon_i)$ is not irreducible it must contain a maximal weight vector
which, in view of Proposition \ref{Prop}, must have weight of the form $\Lambda + \varepsilon_r, r \neq i$. But then this would imply that
\begin{equation*}
\chi_{\Lambda + \varepsilon_i} (I_2) = \chi_{\Lambda + \varepsilon_r} (I_2), ~\hbox{for some}~ r \neq i
\end{equation*}
in contradiction to our basic assumption (\ref{eq:VTVstar}).
Hence it follows that each $L$-module
\begin{equation*}
V(\Lambda + \varepsilon_i) \subseteq U(L_-) V_0 \otimes V_0(\Lambda) \equiv W
\end{equation*}
is irreducible. Thus we have a decomposition into irreducible $L$-modules
\begin{equation}
\label{eq:VTV4}
W 
= \bigoplus^m_{i=1} V(\Lambda + \varepsilon_i),
\end{equation}
induced by the decomposition (\ref{eq:VTV1a}). This then gives an $L_0$-module decomposition
\begin{equation}
\label{eq:VTV5}
V \otimes V(\Lambda) = U(L_-) V_1 \otimes V_0(\Lambda) + W.
\end{equation}

We observe that the form $\left\langle~,~\right\rangle$ on $V \otimes V(\Lambda)$,
restricted to each irreducible $L$-module $V(\Lambda + \varepsilon_i)$ is necessarily
non-degenerate and that the decomposition (\ref{eq:VTV4}) is orthogonal. In view of
(\ref{eq:VTV5}) it follows that the form $\left\langle~,~\right\rangle$ restricted to $W$
is non-degenerate also. We let $P_W$ be the (self-adjoint) projection onto the submodule
(\ref{eq:VTV4}) ~(which thus intertwines the action of $L$). This then gives an orthogonal
decomposition of $L$-modules
\begin{equation}
\label{eq:VTV5prime}
V \otimes V(\Lambda) = W \oplus W^\bot
\end{equation}
where 
\begin{align*}
W^\bot &= (1 - P_W) V \otimes V(\Lambda)\\
&\stackrel{(\ref{eq:VTV5})}{=} (1-P_W)U(L_-)V_1 \otimes V_0(\Lambda)\\
&= U(L_-)(1-P_W)V_1 \otimes V_0(\Lambda).
\end{align*}
We observe that
\begin{align*}
&L_+ (1-P_W) V_1 \otimes V_0(\Lambda)\\
&= (1-P_W)L_+ [V_1 \otimes V_0(\Lambda)]\\
&=(1-P_W) V_0 \otimes V_0(\Lambda)\\
&\subseteq (1-P_W) W = (0)
\end{align*}
so that the decomposition of equation (\ref{eq:VTV1b}) gives 
\begin{equation*}
(1-P_W) V_1 \otimes V_0(\Lambda) = \bigoplus_\mu (1-P_W) V_0 (\Lambda + \delta_\mu)
\end{equation*}
where $L_+ (1-P_W) V_0 (\Lambda + \delta_\mu) = (0)$.

It thus follows that if non-zero, $(1 - P_W) V_0 (\Lambda + \delta_\mu)$ cyclically
generates an indecomposable $L$-module with highest weight $\Lambda + \delta_\mu$. As
before, under our basic assumption (\ref{eq:VTVstar}), this module is in fact irreducible
giving a decomposition of irreducible $L$-modules
\begin{equation}
\label{eq:VTV7}
W^\bot = \bigoplus_\mu V(\Lambda + \delta_\mu).
\end{equation}
The proof of the theorem then follows from the decompositions (\ref{eq:VTV4}),(\ref{eq:VTV5prime}) and (\ref{eq:VTV7}).

It is worth remarking that if we assume that the roots $\alpha_r$, $1\leq r \leq m+n$, of
the characteristic identities are distinct, or equivalently that the numbers
\begin{equation*}
\chi_{\Lambda - \varepsilon_r} (I_2)~~,~ 1 \leq r \leq m+n
\end{equation*}
are distinct, we may similarly prove that:
\begin{thm} 
$V^* \otimes V(\Lambda)$ is completely reducible with $L$-maximal weight vectors of the
form $\Lambda - \varepsilon_r$ $(1\leq r \leq m+n)$, each occurring at most once. 
\end{thm}


\section*{Appendix C}

Here we will define two $gl(m|n)$ invariants that play a central role in
determining reduced matrix elements and reduced Wigner coefficients. We will show that
these invariants are indeed subalgebra invariants by expressing them solely in terms of
the Casimir invariants. 

Here we consider the characteristic matrices ${\cal A}$ and ${\cal B}$ defined in equations
(\ref{equdefx}) and (\ref{equdefy}) respectively. Powers of these matrices are defined recursively 
as given in equations
(\ref{powerx}) and (\ref{powery}) respectively.
By using induction and the $gl(m|n+1)$ commutation relations we obtain
\begin{align}
\label{eq:PowerCOMRELATION}
\left[ E^p_q , \left({\cal B}^k\right)^r_s \right] = (-1)^{(p) + (q)}
\delta^r_q \left({\cal B}^k\right)^p_s - (-1)^{((p) + (q))((r) +
(s))} \delta^p_s \left({\cal B}^k\right)^r_q.
\end{align}
For readability, throughout this appendix the '$+$' symbol will be used to denote the odd value $m+n+1$. A special
case of the above identity is then
\begin{align}
\label{eq:PowerCOMRELATION2}
\left[ E^p_q , \left({\cal B}^k\right)^+_+ \right] = -(-1)^{(p)} \delta^+_q
\left({\cal B}^k\right)^p_+ - \delta^p_+ \left({\cal B}^k\right)^+_q
\end{align}
or equivalently
\begin{align}
\label{eq:PowerCOMRELATION3}
\left[ {\cal B}^p_q , \left({\cal B}^k\right)^+_+ \right] = (-1)^{(p)} \left[
E^p_q , \left({\cal B}^k\right)^+_+ \right] = - \delta^+_q \left({\cal
B}^k\right)^p_+ + \delta^p_+ \left({\cal B}^k\right)^+_q.
\end{align}

In what follows, we also use the index notation $\dot{p}$ to denote an index ranging from $1$ to $m+n$
only.

The invariants under consideration in this appendix are defined as
\begin{align}
\tau_k &= ({\cal B}^k)^+_+\nn\\
\sigma_k &= {\cal B}^+_{\dot{p}} ({\cal A}^k)^{\dot{p}}_{\dot{q}} {\cal B}^{\dot{q}}_+.\nn
\end{align}

\begin{prop} \label{PropBsCommute}
The set $\{\tau_1, \tau_2, \tau_3 ,... \}$ is a set of commuting operators. That is, 
\begin{align}
\label{eq:BsCOMMUTE}
[\tau_\ell,\tau_k] = 0 ~~\forall \ell,k
\end{align}
\end{prop}
\textbf{Proof:}
We proceed by induction. For $\ell=1$, equation (\ref{eq:BsCOMMUTE}) is seen to be valid
immediately by applying the commutation relation given in equation
(\ref{eq:PowerCOMRELATION3}).
\begin{align}
[\tau_1,\tau_k] &= \left[{\cal B}^+_+, ({\cal B}^k)^+_+\right]\nn\\
&= -\delta^+_+ \left({\cal B}^k\right)^+_+ + \delta^+_+ \left({\cal B}^k\right)^+_+\nn\\
&= 0 ~~ ~~\forall k.
\end{align}
For $\ell > 1$ we have
\begin{align}
[\tau_\ell,\tau_k] &= \left[({\cal B}^\ell)^+_+, ({\cal B}^k)^+_+\right]\nn\\
&= ({\cal B}^{\ell-1})^+_q \left[{\cal B}^q_+, ({\cal B}^k)^+_+\right] + ({\cal
B}^{\ell-2})^+_q \left[ {\cal B}^q_r, ({\cal B}^k)^+_+ \right] {\cal
B}^r_+ + ({\cal B}^{\ell-3})^+_q \left[{\cal B}^q_r, ({\cal B}^k)^+_+\right] ({\cal
B}^2)^r_+ \nn\\
&~~~~ + ... + \left[{\cal B}^+_p,({\cal B}^k)^+_+\right] ({\cal B}^{\ell-1})^p_+ \nn\\
&= ({\cal B}^{\ell-1})^+_q \left[{\cal B}^q_+, ({\cal B}^k)^+_+\right] +
\left[{\cal B}^+_p,({\cal B}^k)^+_+\right]
({\cal B}^{\ell-1})^p_+ + \sum_{j=1}^{\ell-2} ({\cal B}^{\ell-j-1})^+_q \left[ {\cal
B}^q_r, ({\cal B}^k)^+_+
\right] ({\cal B}^j)^r_+ .\nn
\end{align}
By use of equation (\ref{eq:PowerCOMRELATION3}) to evaluate the commutators, one can
readily show that
\begin{align}
\sum_{j=1}^{\ell-2} ({\cal B}^{\ell-j-1})^+_q \left[ {\cal B}^q_r, ({\cal B}^k)^+_+ \right]
({\cal B}^j)^r_+ 
&= \sum_{j=1}^{\ell-2} \left[ \tau_j,\tau_{k+\ell-j-1}  \right]\nn
\end{align}
and also
\begin{align}
({\cal B}^{\ell-1})^+_q \left[{\cal B}^q_+, ({\cal B}^k)^+_+\right] + \left[{\cal
B}^+_p,({\cal B}^k)^+_+\right]
({\cal B}^{\ell-1})^p_+ &= ({\cal B}^{\ell-1})^+_q \left( -({\cal B}^k)^q_+ + \delta^q_+ ({\cal B}^k)^+_+
\right)\nn\\ &~~~+ \left(-\delta^+_p ({\cal B}^k)^+_+  + ({\cal B}^k)^+_p   \right)
({\cal B}^{\ell-1})^p_+\nn\\
&= - \tau_{\ell+k-1} + \tau_{\ell-1}\tau_k - \tau_k \tau_{\ell-1} + \tau_{\ell+k-1}\nn\\
& = [\tau_{\ell-1},\tau_k].\nn
\end{align}
We have shown that $[\tau_\ell,\tau_k]$ can be written in terms of commutators with lower
order $\tau$ invariants on the LHS. That is,
\begin{align}
[\tau_\ell,\tau_k] = \sum_{j=1}^{\ell-1} \left[ \tau_j,\tau_{k+\ell-j-1}  \right] \nn
\end{align}
so by induction we have $[\tau_\ell,\tau_k] = 0$ since $[\tau_1,\tau_k] = 0 ~\forall k$.
$\blacksquare$ 

The next three propositions will be required when expressing $\tau_k$ in terms of the
Casimir invariants. Recall the Casimir invariants $\hat{I}_k$ and $I_k$ are defined in
equations (\ref{casop1}) and (\ref{casop2}) respectively. Using the commutation relations
(\ref{eq:PowerCOMRELATION}) the following result can be established:
\begin{prop} \label{PropUpperLower}
\begin{align}
\label{eq:UpperLowerCommutator}
\sum_{p = 1}^{m+n+1}(-1)^{(p)} \left[ ({\cal B}^\ell)^p_+,({\cal B}^k)^+_p \right] =
\sum_{i=0}^{\ell-1} \left( \hat{I}_i \tau_{\ell+k-1-i} - \hat{I}_{\ell+k-1-i}\tau_i \right).
\end{align}
\end{prop}

\textbf{Definition.}
The \textit{order} of the term 
\begin{align}
(-1)^{(p)}({\cal B}^k)^p_+ \tau_{s_1} \tau_{s_2} \ldots \tau_{s_j} ({\cal B}^\ell)^+_p
\nn\end{align}
is defined to be the sum of the powers of the ${\cal B}$'s within the term which in the above case is
\begin{align}
k+\ell + \sum_{i=1}^j s_i.
\nn\end{align}

It is important to note that for sufficiently low orders, this term will degrade to two possible cases. The first case is when $k=0$ which gives 
\begin{align}
\sum_{p}(-1)^{(p)}({\cal B}^0)^p_+ \tau_{s_1} \tau_{s_2} \ldots \tau_{s_j} ({\cal
B}^\ell)^+_p &= \sum_{p}(-1)^{(p)} \delta^p_+ \tau_{s_1} \tau_{s_2} \ldots \tau_{s_j} ({\cal B}^\ell)^+_p\nn\\
&= -\tau_{s_1} \tau_{s_2} \ldots \tau_{s_j} ({\cal B}^\ell)^+_+\nn\\
&= -\tau_{s_1} \tau_{s_2} \ldots \tau_{s_j} \tau_\ell.\nn
\end{align}
The second case is $s_i = 0 ~~ \forall i$ where proposition \ref{PropUpperLower} gives
\begin{align}
\sum_p(-1)^{(p)}({\cal B}^k)^p_+ \tau_0 \tau_0 \ldots \tau_0 ({\cal
B}^\ell)^+_p &= \sum_p(-1)^{(p)}({\cal B}^k)^p_+ ({\cal B}^\ell)^+_p\nn\\
&= \sum_p(-1)^{(p)} [({\cal B}^k)^p_+,({\cal B}^\ell)^+_p] + \tau_{k+\ell}\nn\\
&= \sum_{i=0}^{k-1} \left(\hat{I}_i \tau_{k+\ell-i-1} - \hat{I}_{k+\ell-i-1} \tau_i \right) +
\tau_{k+\ell}.\nn
\end{align}

By repeated use of the graded commutation relations we have:
\begin{prop} 
\label{PropLBBBUFinal}
A general term of the form
\begin{align} 
\sum_{p=1}^{m+n+1} (-1)^{(p)}({\cal B}^k)^p_+ \tau_{s_1} \tau_{s_2}\ldots\tau_{s_j} ({\cal B}^\ell)^+_p 
\nn\end{align}
which is of order
\begin{align}
M = k+\ell + \sum_{i=1}^j s_i
\nn\end{align}
can be written as a series of terms of the form
\begin{align}
\tau_a \tau_b \ldots \tau_c  ~~~\hbox{and}~~~\hat{I}_a \tau_b
\nn\end{align}
(and products of them) where the order of each term (and therefore the order of each $\tau$ and $I$ within a term) is strictly less than $M$.
\end{prop}

We are now in a position to state the two main theorems of this appendix.

\begin{thm} \label{TheoremBandIB}
The invariant $\tau_k$ can be written as the sum of $I_j - \hat{I}_j$ and a series of terms of the form
\begin{align}
\tau_a \tau_b \ldots \tau_c  ~~~\hbox{and}~~~\hat{I}_a \tau_b
\nn\end{align}
(and products of them) where the order of each term (and therefore the order of each $\tau$ and $I$ within a term) is strictly less than $k$.
\end{thm}
\textbf{Proof:}
Firstly, we consider the difference of the Casimir invariants $\hat{I}_k$ and $I_k$. By definition, we have
\begin{align}
\hat{I}_k = (-1)^{(p)}({\cal B}^k)^p_p =  (-1)^{(p)}{\cal B}^p_{q_1} {\cal
B}^{q_1}_{q_2} {\cal B}^{q_2}_{q_3}\ldots{\cal B}^{q_{k-1}}_{p} \nn
\end{align}
and 
\begin{align}
I_k &= (-1)^{(p)}({\cal A}^k)^p_p ~~~~1 \leq p \leq m+n\nn\\
&= (-1)^{(p)}(1-\delta^p_+)({\cal B}^k)^p_p ~~~~1 \leq p \leq
m+n+1\nn\\
&=
(-1)^{(p)}(1-\delta^p_+)(1-\delta^{q_1}_+)(1-\delta^{q_2}_+)\ldots(1-\delta^{q_{k-1}}_+)
{\cal B}^p_{q_1} {\cal B}^{q_1}_{q_2} {\cal
B}^{q_2}_{q_3}\ldots{\cal B}^{q_{k-1}}_{p}\nn\\
&=
\delta^p_+(1-\delta^{q_1}_+)(1-\delta^{q_2}_+)\ldots(1-\delta^{q_{k-1}}_+)
{\cal B}^p_{q_1} {\cal B}^{q_1}_{q_2} {\cal
B}^{q_2}_{q_3}\ldots{\cal B}^{q_{k-1}}_{p}\nn\\
&~~+
(-1)^{(p)}(1-\delta^{q_1}_+)(1-\delta^{q_2}_+)\ldots(1-\delta^{q_{k-1}}_+)
{\cal B}^p_{q_1} {\cal B}^{q_1}_{q_2} {\cal
B}^{q_2}_{q_3}\ldots{\cal B}^{q_{k-1}}_{p}.
\nn\end{align}
Now we note that
\begin{align}
\delta^p_+ ({\cal B}^p_{q_1} {\cal B}^{q_1}_{q_2} {\cal
B}^{q_2}_{q_3}\ldots{\cal B}^{q_{k-1}}_{p} ) &= \tau_k\nn\\
\delta^p_+ \delta^{q_i}_+  ({\cal B}^p_{q_1} {\cal B}^{q_1}_{q_2} {\cal
B}^{q_2}_{q_3}\ldots{\cal B}^{q_{k-1}}_{p})&= {\cal B}^+_{q_1} \ldots{\cal B}^{q_{i-1}}_+ {\cal
B}^+_{q_{i+1}}\ldots{\cal B}^{q_{k-1}}_+ \nn\\
&= \tau_i \tau_{k-i}\nn\\
\delta^p_+ \delta^{q_i}_+ \delta^{q_j}_+ ({\cal B}^p_{q_1} {\cal B}^{q_1}_{q_2} {\cal
B}^{q_2}_{q_3}\ldots{\cal B}^{q_{k-1}}_{p})&= {\cal B}^+_{q_1} \ldots{\cal B}^{q_{i-1}}_+ {\cal
B}^+_{q_{i+1}}\ldots{\cal B}^{q_{j-1}}_+ {\cal B}^+_{q_{j+1}}\ldots{\cal B}^{q_{k-1}}_+ \nn\\
&= \tau_i \tau_{j-i} \tau_{k-j} \hbox{~for } i < j \nn\\
&\cdots\nn
\end{align}
and observe that in each case the terms are of order $k$ (since the subscripts within each term sum to $k$).
Similarly we have
\begin{align}
(-1)^{(p)}({\cal B}^p_{q_1} {\cal B}^{q_1}_{q_2} {\cal
B}^{q_2}_{q_3}\ldots{\cal B}^{q_{k-1}}_{p} ) &= \hat{I}_k\nn\\
(-1)^{(p)}\delta^{q_i}_+  ({\cal B}^p_{q_1} {\cal B}^{q_1}_{q_2}
{\cal B}^{q_2}_{q_3}\ldots{\cal B}^{q_{k-1}}_{p})&=  (-1)^{(p)}
{\cal B}^p_{q_1}\ldots{\cal B}^{q_{i-1}}_+ {\cal
B}^+_{q_{i+1}}\ldots{\cal B}^{q_{k-1}}_p \nn\\
&= (-1)^{(p)}({\cal B}^i)^p_+ ({\cal B}^{k-i})^+_p\nn\\
(-1)^{(p)}\delta^{q_i}_+ \delta^{q_j}_+ ({\cal B}^p_{q_1} {\cal
B}^{q_1}_{q_2} {\cal B}^{q_2}_{q_3}\ldots{\cal
B}^{q_{k-1}}_{p})&= (-1)^{(p)} {\cal B}^p_{q_1} \ldots{\cal
B}^{q_{i-1}}_+ {\cal B}^+_{q_{i+1}}\ldots{\cal B}^{q_{j-1}}_+ {\cal
B}^+_{q_{j+1}}\ldots{\cal B}^{q_{k-1}}_p \nn\\
&= (-1)^{(p)} ({\cal B}^i)^p_+ \tau_{j-i} ({\cal B}^{k-j})^+_p \hbox{~for }
i < j \nn\\
&\ \cdots \nn
\end{align}
where each term is also of order $k$.
By defining the summation symbol $\sum_{(k)}$ to be the sum of all terms such that the
powers of ${\cal B}$ within each term are positive and sum to $k$ (implying that each term
is of order $k$) we can write $I_k$ as
\begin{align}
I_k &=  \tau_k - \sum_{(k)} \tau_a \tau_b + \sum_{(k)}\tau_a \tau_b \tau_c -
\sum_{(k)}\tau_a \tau_b \tau_c \tau_d + ... \nn\\
&~~ + \hat{I}_k - (-1)^{(p)} \sum_{(k)} ({\cal B}^i)^p_+ ({\cal B}^j)^+_p + (-1)^{(p)}
\sum_{(k)}({\cal B}^i)^p_+ \tau_j ({\cal B}^\ell)^+_p - (-1)^{(p)}
\sum_{(k)}({\cal B}^i)^p_+ \tau_j
\tau_\ell ({\cal B}^h)^+_p + \ldots\nn
\end{align}
Rearranging gives
\begin{align}
\label{eq:BkExpansion}
\tau_k &= I_k - \hat{I}_k  + \sum_{(k)} \tau_a \tau_b - \sum_{(k)}\tau_a \tau_b \tau_c +
\sum_{(k)}\tau_a \tau_b \tau_c \tau_d - ... \nn\\
&~~~ + (-1)^{(p)}\sum_{(k)} ({\cal B}^i)^p_+ ({\cal B}^j)^+_p -
(-1)^{(p)}\sum_{(k)}({\cal B}^i)^p_+ \tau_j ({\cal B}^\ell)^+_p +
(-1)^{(p)}\sum_{(k)}({\cal B}^i)^p_+ \tau_j \tau_\ell ({\cal B}^h)^+_p - ...
\nn\end{align}
which together with proposition~\ref{PropLBBBUFinal} completes the proof.
$\blacksquare$ 

\begin{thm} 
The invariants $\tau_k,\sigma_\ell$ can be expressed in terms of products of Casimir
invariants $I_j$ and $\hat{I_j}$ where $j\leq k$ and $j\leq \ell+2$.
\end{thm}
\textbf{Proof:}
For $\tau_k$
the proof is obtained by applying Theorem \ref{TheoremBandIB} recursively and
observing that $\tau_0 = 1$ and $\tau_1 = I_1 - \hat{I}_1$ (since $\tau_1 = {\cal B}^+_+ = - E^+_+$).

For the $\sigma_\ell$ we have
\begin{align}
\sigma_\ell
&= {\cal B}^+_{\dot{p}} ({\cal A}^\ell)^{\dot{p}}_{\dot{q}} {\cal B}^{\dot{q}}_+ ~~~(1 \leq
\dot{p},\dot{q} \leq m+n)\nn\\
&= (1-\delta^{q_1}_+)(1-\delta^{q_2}_+)...(1-\delta^{q_{\ell+1}}_+) {\cal B}^+_{q_1}
{\cal B}^{q_1}_{q_2} {\cal B}^{q_2}_{q_3}...{\cal B}^{q_{\ell+1}}_+ 
~~~(1 \leq q_i \leq m+n+1)\nn\\
&= \left(1 - \sum_i \delta^{q_i}_+ + \sum_{i,j~i \neq j}  \delta^{q_i}_+
\delta^{q_j}_+ - \sum_{i,j,k~i \neq j \neq k} \delta^{q_i}_+
\delta^{q_j}_+ \delta^{q_k}_+ + ...\right)  {\cal B}^+_{q_1} {\cal
B}^{q_1}_{q_2} {\cal B}^{q_2}_{q_3}...{\cal B}^{q_{\ell+1}}_+ \nn\\
&= \tau_{\ell+2} - \sum_{i=1}^{\ell+1} \tau_i \tau_{\ell+2-i} + \sum_{(\ell+2)} \tau_a \tau_b \tau_c -
\sum_{(\ell+2)} \tau_a \tau_b \tau_c \tau_d + ...
\end{align}
where again the summation symbol $\sum_{(\ell+2)}$ is defined to be the sum of all terms
such that the powers of ${\cal B}$ within each term are positive and sum to $\ell+2$.
$\blacksquare$ 

As an example, we give $\sigma_2$ in terms of $\tau$'s as 
\begin{align}
\sigma_2 &= \tau_4 - \tau_1\tau_3 - \tau_2 \tau_2 - \tau_3 \tau_1 
+ \tau_1 \tau_1 \tau_2 + \tau_1 \tau_2 \tau_1 + \tau_2 \tau_1 \tau_1
- (\tau_1)^4. \nn
\end{align}

%
%
%
%
%
%
%
%

\end{document}